\documentclass{aa}  
\usepackage{natbib}
\bibpunct{(}{)}{;}{a}{}{,} 
\usepackage{graphicx}
\usepackage{txfonts}
\usepackage{hyperref}
\begin{document} 
\newcommand{\km}        {\ensuremath{\mathrm{\ km}}}
\newcommand{\s}         {\ensuremath{\mathrm{\ s}}}
\newcommand{\kms}       {\ensuremath{\mathrm{\km\s^{-1}}}}
\newcommand{\HI}        {\ensuremath{\rm H\,{\scriptstyle I}}\,}
\newcommand{\Hmol}        {\ensuremath{\rm H_{\scriptstyle 2}}\,}
\newcommand{\Otwo}        {O$_2$}
\newcommand{\Othree}        {O$_3$}
\newcommand{\HHO}        {H$_2$O}
\newcommand{\KI}        {\ensuremath{\rm K\,{\scriptstyle I}}\,}
\newcommand{\NaI}        {\ensuremath{\rm Na\,{\scriptstyle I}}\,}
\newcommand{\CaII}        {\ensuremath{\rm Ca\,{\scriptstyle II}}\,}
\newcommand{\CaI}        {\ensuremath{\rm Ca\,{\scriptstyle I}}\,}
\newcommand{\Ca}        {\ensuremath{\rm Ca}\,}
\newcommand{\Na}        {\ensuremath{\rm Na}\,}
\newcommand{\NaD}        {\ensuremath{\rm Na\,{\scriptstyle I\, D}}\,}
\newcommand{\ed}        {\b}

\title{Mapping diffuse interstellar bands in the local ISM on small scales via MUSE 3D spectroscopy}

   \subtitle{A pilot study based on globular cluster NGC\,6397}

   \author{Martin Wendt\inst{1,2}
          \and
          Tim-Oliver Husser\inst{3}
          \and
          Sebastian Kamann\inst{3}
          \and
          Ana Monreal-Ibero\inst{4,5,6}
          \and
          Philipp Richter\inst{1,2}
         \and
          Jarle Brinchmann\inst{7}
          \and
          Stefan Dreizler\inst{3}
        \and
          Peter M.\ Weilbacher\inst{2}
          \and
          Lutz Wisotzki\inst{2}}

      \institute{Institut für Physik und Astronomie, Universität Potsdam,Karl-Liebknecht-Str. 24/25,  
14476 Golm, Germany\\
              \email{mwendt@astro.physik.uni-potsdam.de}
         \and
          Leibniz-Institut für Astrophysik Potsdam (AIP), An der Sternwarte 16, 14482 Potsdam, 
Germany
          \and
          Institut für Astrophysik, Georg-August-Universität Göttingen, Friedrich-Hund-Platz 1, 37077 
Göttingen, Germany
\and
          GEPI, Observatoire de Paris, PSL Research University, CNRS, Universit\'e Paris-Diderot, Sorbonne Paris Cit\'e, Place Jules Janssen, 92195 Meudon, France
\and
Instituto de Astrof\'isica de Canarias (IAC), E-38205 La Laguna, Tenerife, Spain
\and
Universidad de La Laguna, Dpto. Astrof\'isica, E-38206 La Laguna, Tenerife, Spain
    \and 
       Leiden Observatory, Leiden University, PO Box 9513, 2300 RA Leiden, The Netherlands }

   \date{Received \today; accepted \today}
\abstract
   {
We map the interstellar medium (ISM) including the diffuse interstellar bands 
(DIBs) 
   in absorption toward the globular cluster NGC\,6397 using VLT/MUSE. 
   Assuming the absorbers are located at the rim of the Local Bubble we trace structures 
on the order
   of mpc (milliparsec, a few thousand AU). 
  } 
   {
{We aimed to demonstrate the feasibility to map variations of DIBs on small scales with MUSE.} The sightlines
defined by binned stellar spectra are  separated by only a few arcseconds and we probe the absorption within a physically connected region. 
   }  
   {
This analysis utilized the fitting 
residuals of individual stellar spectra of NGC\,6397 member stars by \citet{paperi} and \citet{paperii}
   and analyzed  lines from neutral species and several DIBs in Voronoi-binned composite spectra with 
high signal-to-noise ratio (S/N).
}
   {
   {This pilot study demonstrates the power of MUSE for mapping 
the local ISM on very small scales which provides a new window for ISM observations.}

We detect small scale variations { in \NaI and \KI as well as in several DIBs  within few} 
arcseconds, or mpc  with regard to the Local Bubble. We verify the suitability of the MUSE 3D spectrograph
for such measurements and gain new insights by probing a single physical absorber with multiple sight lines.
}
   {}


   \maketitle
%

\section{Introduction}
Systematic studies of the interstellar medium (ISM) are of prime
importance to understanding the life-cycle of baryons in the Universe
and the evolution of galaxies. Because the ISM spans several orders
of magnitudes in gas densities and temperatures and is structured
down to AU scales, our understanding of the ISM is still quite
limited.

Among the various different methods of studying the ISM, absorption
spectroscopy in the ultraviolet (UV) and optical regime has become
a particularly powerful approach to explore the ISM's chemical
composition, kinematics, and spatial structure.
Not long after the first detection of interstellar absorption
lines in the Ca H \& K lines \citep{Hartmann1904} a number of broad
optical absorption features of unknown origin were discovered
\citep{Heger1922}. These features were given the name
diffuse interstellar bands (DIBs). Today, more than
400 DIBs have been identified in the Milky Way's ISM
(\citealt{Herbig1995}; \citealt{Sarre2006}) and they  have also been observed
in other galaxies (e.g., in the Magellanic Clouds; \citealt{Walker1963};
\citealt{Ehrenfreund2002} or even beyond the local group by \citealt{ana15b}).
 Even almost 100 years after their
first detection, however, the exact origin of the DIBs (i.e., their carriers)
is largely uncertain. One preferred scenario is, that DIBs represent
carbon-based molecular structures and possibly are related to
polycyclic aromatic hydrocarbons (PAHs; e.g., \citealt{Crawford1985}, \citealt{Cox2011}).
 The strengths of the various DIBs show a
rather complex behavior when compared to other tracers of
interstellar gas and dust particles, such as the color excess
$E(B-V)$ and the equivalent width of neutral and molecular
species (e.g., Na\,{\sc i}, CN). This indicates that the
DIB carriers are different from those of interstellar dust particles
and simple molecules (e.g., \citealt{Friedman11}).
Despite the uncertain origin, DIBs can be used as a diagnostic tool to
study the radiation field and gas temperature in the ISM \citep{Cami1997,vos}.
And, if multiple sightlines at small angular separations are used, DIBs do
serve as tracer species to study the small-scale structure in the ISM
(\citealt{loon09}, \citealt{loon13}; \citealt{Cordiner}).
The large number of different DIBs and their commonness render the nature of DIBs a pressing puzzle.

\citet{Andrews2001} and \citet{loon09} present indications for small scale structures in the ISM in individual
sightlines toward globular cluster stars on parsec scale, while \citet{a7} and \citet{Boisse2013} tested for spatial
variations on even smaller scales utilizing the proper motion of stars and the implied drift of the line of sight through the foreground gas.

The existence of small regions with comparably high density within the
diffuse ISM has important implications for interstellar chemistry. Their detection might also provide a potential method
for obtaining new information on the physical conditions, and a possible solution to some
 of the challenges which exist in reproducing the abundances of molecules
observed along lines of sight through the diffuse ISM \citep{a7}.
\citet{Andrews2001} point out that fluctuations in ionization equilibrium
and not just total column density may be responsible for the observed variations in the ISM on small scales.
It appears to be essential to take the spatial information into account for the 
quest to understand the nature of the DIBs as well as their carriers.
While, for example, \citet{ana15} trace the extinction on the plane of the sky and deepen
the knowledge on the DIB correlations with dust, works like 
\citet{loon15} provide further strong evidence
 that the origin of the DIBs is manifold and they do indeed show structures on small scales,
which help to differentiate between them.  
{ Despite the number of detections, the nature of these  
structures and their ubiquity is still a subject of study. It
has not always been clear whether they reflect the variation in
\HI column density, or whether they are caused by changes in
the physical conditions of the gas over small scales. 
}

In this paper we take advantage of the manifold capabilities of the
Multi Unit Spectroscopic Explorer (MUSE) installed at the
ESO Very Large Telescope (VLT) to study DIBs and optical absorption
lines in the local ISM at very small angular scales in the direction
of the globular cluster NGC\,6397 in the broad wavelength range between 4650\AA\, and 9300\AA.
The MUSE 3D spectrograph provides a respectable FOV of $1' \times 1'$ and a spatial 
sampling of  $0.2'' \times 0.2''$ which enables us to construct equivalent width maps of 
DIBs and of their correlation with respect to each other or with the observed ISM 
features for a physically connected region.
{ DIBs that are related to the same carrier should reveal themselves in similar 2D maps.
With the high spatial resolution we trace the correlation between DIBs within connected regions which
can disclose the preferred environment of individual DIBs.}

In Sect. \ref{sec:data} we describe the MUSE observations and several ancillary data taken with 
other instruments. Section \ref{sec:extraction} depicts the applied method to obtain information on the ISM absorption
features from the MUSE data cubes and the results are presented in Sect. \ref{sec:results}.
We discuss the new findings and its implications in Sect. \ref{sec:discussion}.
Further plots and details on the fitting procedure can be found in the Appendix.
Throughout the paper the wavelengths $\lambda$ is given in \AA, e.g., \KI 7664, DIB 5780, and equivalent
widths in m\AA, unless noted otherwise.

\section{The data}
\label{sec:data}
NGC\,6397 is the second closest globular cluster with a distance of merely 2.3 kpc
and a reddening of E(B-V) = 0.18 \citep{harris}.
At a galactic latitude of -11.06$^{\circ}$ and longitude of 
+338.17$^{\circ}$,
the globular cluster (GC) sits below the Galactic disk at about 6 kpc distance to the Galactic 
center.
The GC has a mean radial velocity of $v_{\mathrm{rad}}=17.84\pm0.07$\kms
 (see \citealt{paperi}, hereafter referred to as Paper I).
 NGC\,6397 is often considered to be one of the 20 globulars (at least) in our Milky Way
 Galaxy that have undergone a core collapse.
 Core collapse is when a globular's core has contracted to a very dense stellar agglomeration
 (see \citealt{paperii}, hereafter referred to as Paper II).

\subsection{The MUSE data}
 NGC\,6397 was observed during MUSE commissioning, lasting from July 26nd to August 3rd,
2014. 
23 MUSE pointings were arranged in a $5 \times 5$ mosaic with two missing frames, covering roughly $300' \times 
300'$ on the sky.
The exposure times ranged from 25s to 60s per individual observation, accumulating about 4 min per pointing
on average. This limitation was set to avoid
saturation effects of the brightest cluster giants as the original studies in Papers I \& II focused on the stars themselves.
 In total, we obtained 127 exposures with a total integration time of 95 min. Including
overheads, the observations took about 6 h. While the seeing varied below $1''$, the central pointings
were recorded at a seeing of $0.6''$.
The  data  reduction  was  done  using  the  official  MUSE pipeline (in versions 0.18.1, 0.92, and 1.0)
 described in \citet {weilbacher12}. See Paper I for detailed information on the MUSE observations of NGC\,6397.
{The resolving power of the MUSE instrument ranges from $\sim 2000$ at 4650\AA\, to $\sim 4000$ at 9300\AA.
The data consists of roughly 19,000 spectra 
 of $\sim$ 12,000 individual 
stars. Even at a resolution of $\approx 100$ \kms, most of the strong DIBs are resolved, due to their eponymous broad profiles.
However, we are strictly limited by the resolution of our data with regard to resolving individual velocity components in the narrow
{ absorption lines of neutral species such as \KI\, or the \NaD\, doublet.}
}

{
The observed field is strongly dominated by cluster member stars. We expected comparably few foreground stars toward
the position of NGC\,6397 below the Galactic disk. For an adjacent field of the same angular size, the GAIA archive of Data Release 1 contains 
$\approx$ 100 stars brighter than 18th magnitude (see \citealt{gaiadr1}, \citealt{Fabricius2016}).
18th magnitude stars correspond to a S/N of approximately ten which we used as lower threshold in our sample.
As will be discussed in Sect. \ref{sec:location} there is strong evidence that the absorbing cloud is nearby but 
even if those $\approx$ 100 stars were distributed highly inhomogeneously across the field, they would not impact the
results. Each spatial measurement was based on more than 300 stars (see Sect. \ref{sec:voronoi}).
} 

\begin{figure}[hb!]
   \centering
    \includegraphics[width=0.5\textwidth]{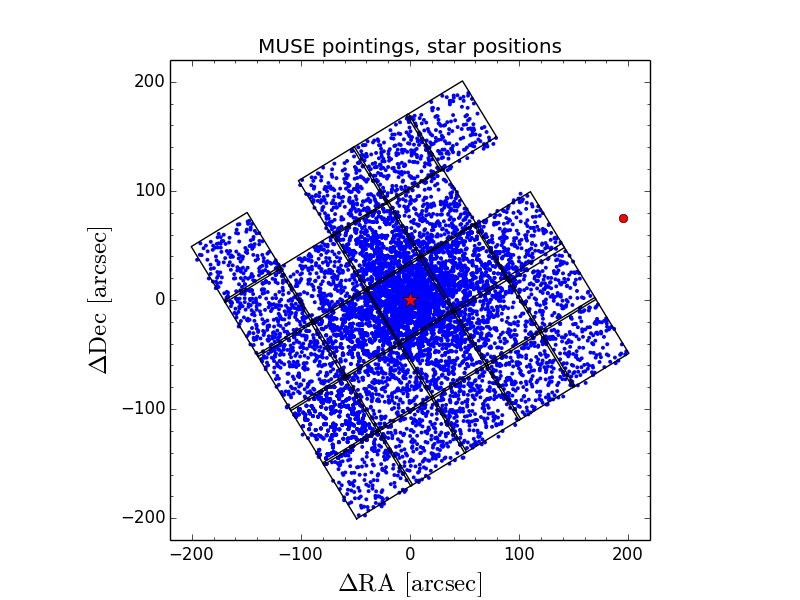}
    \caption{Area covered by our observations: {\it black:} MUSE pointings, {\it blue:} individual stellar positions, {\it red circle:} Single High resolution UVES pointing
  relative to the center of NGC\,6397 ({\it red star}).}
    \label{fig:pointings}%
    \end{figure}

In Fig. \ref{fig:pointings} we plot all stars relative to the cluster center. The
individual pointings are indicated. 
\subsection{Ancillary data}
   \begin{figure}[h!]
   \centering
   \includegraphics[width=0.45\textwidth]{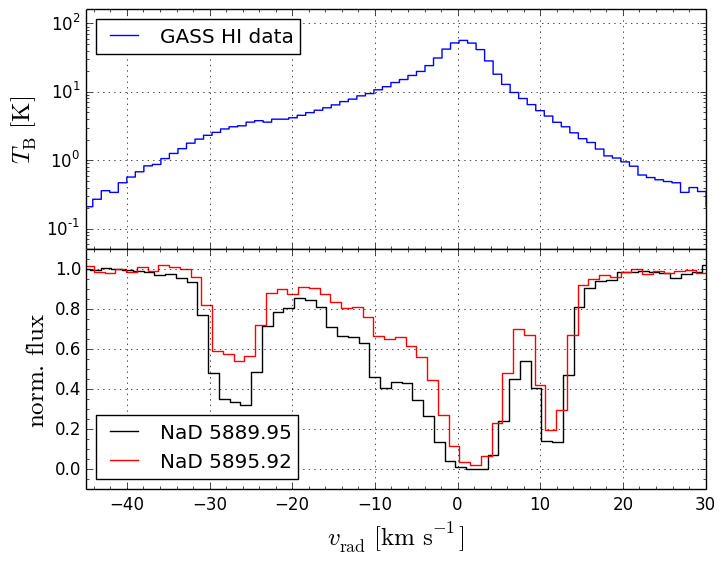}\\
      \includegraphics[width=0.45\textwidth]{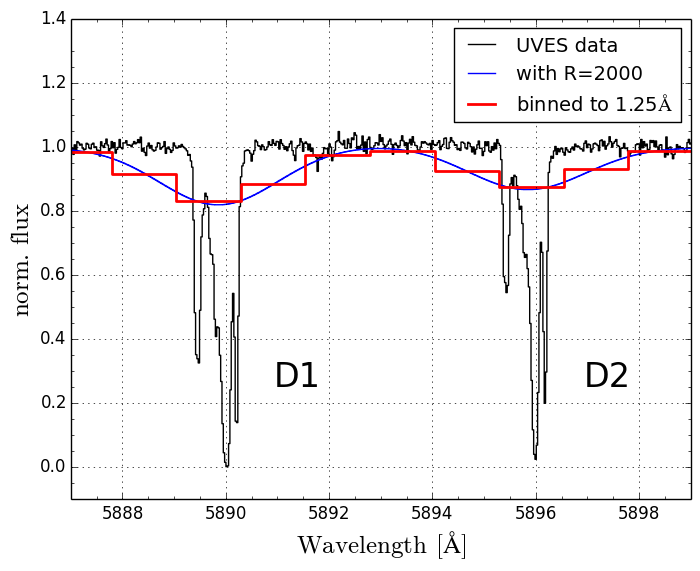}
      \caption{{\it top panel:}UVES spectrum of \NaD doublet + \HI GASS data ({\it blue}),
               {\it bottom panel:}UVES spectrum of \NaD doublet D1 and D2 ({\it black}), convolved to MUSE resolution ({\it blue})
              and binned to default MUSE sampling of 1.25\AA\, ({\it red}). The scale for all data is helio-centric.}
         \label{fig:uves1}
   \end{figure}

\subsubsection{UVES data}
We used a high resolution spectrum from VLT/UVES in the direction of NGC\,6397 of one of its cluster members 
to verify the results we derived from low resolution
MUSE data and to
provide a reference spectrum for interstellar absorption.
A marker in Fig. \ref{fig:pointings}, about 200 arcsecs from the cluster center, reflects the position of
star NGC\,6397-T183. 
{  This star lies outside of our map and also outside of the cluster's 
core region, but since it is relatively close on the sky, its spectrum should give useful indications on the velocity components present in this direction.
We retrieved a high resolution UVES spectrum from the 
ESO archive\footnote{Program: 081.D-0498(A), PI: Hubrig, S.}.}
At a slit width of $0.3''$, the UVES spectrum has a resolution of $R \approx 115,000$ and a S/N of $\sim$ 50.
The UVES data for the nearby B-Type star\footnote{See \citet{t183}.} in the top panel of Fig. \ref{fig:uves1} reveal three components. 
At $v_{\mathrm{helio}}\approx$ -26 \kms, the main saturated component at rest and the stellar component at $\approx$ +12 \kms
              for this individual star. There is a weak fourth component at $\approx$ -10 \kms.
 
{ The wavelength region around the \NaD absorption for this individual cluster star is shown in Fig. \ref{fig:uves1}. A comparison of these data at their original resolution and once degraded to the 
MUSE spectral resolution is illustrated.}
The plot shows the same region of the UVES data in wavelength space. The resulting spectrum
after a convolution with a Gaussian kernel to the MUSE resolution of $R \approx 2,200$ at the given wavelength is also shown, as 
are the 1.25\AA\, sampling of the MUSE data as present in the MUSE pipeline-reduced data cubes. 
{In addition to \NaD, only DIB 5780 and DIB 6283 are covered in the high resolution data (see Sect. \ref{sec:discussion}).}

The analysis presented in this paper is solely based on the MUSE data, with exceptions indicated in the text.
We note that the stellar component (as well as the unresolved telluric absorption component) 
will be removed prior to data analysis as described in  
Sect. \ref{sec:extraction}. The findings based on the UVES data are further discussed in Sect. \ref{sec:uvesmuse}.
\subsubsection{GASS data on \HI}
To further study the gas kinematics of the neutral ISM toward
NGC\,6397 (including the more diffuse warm neutral phase that is
not traced by Na\,{\sc i} absorption) we used publicly available H\,{\sc i} 21cm
data from the Galactic-All Sky Survey (GASS; \citealt{Griffiths2009};
\citealt{Kalberla2010}; see {\tt https://www.astro.uni-bonn.de/hisurvey)}.
The GASS survey was carried out with the 64m radio telescope at Parkes.
{
The spectral
resolution is $0.82$ km\,s$^{-1}$ at an rms of $57$ mK per spectral channel
($\Delta v=0.8$ km\,s$^{-1}$), while the beam size of the data
 is $\sim 16 \arcmin$, thus providing a rough estimate of $N(\HI)$ as the 
effective FWHM of the beam is larger than the MUSE FOVs.}

{ 
If one assumes that the gas is optically thin, the column density of \HI can then be obtained 
from the brightness temperature $T(v)$ per velocity bin from the GASS data:
\begin{equation}
N_{\mathrm{H}} = 1.82 \times 10^{13} \int T(v)\,dv\,\,\,\mathrm{atoms\,cm}^{-2},
\end{equation}
leading to a total column density of $\log[N(\HI)]= 21.1$.
As this is observed in emission, gas beyond NGC\,6397 could contribute to the total budget.
The low galactic latitude of -11.06$^{\circ}$ for NGC\,6397 suggests that all observed 
\HI\,is in the foreground of the GC.}

The 21cm brightness temperature profile toward NGC\,6397 is shown in the top panel in Fig. \ref{fig:uves1}.
The radio data clearly displays the main component at
rest as well as the much weaker blue shifted component at
$v_{\rm helio}=-26$ km\,s$^{-1}$. 

\subsubsection{Extinction data}
The reddening E(B-V) toward NGC\,6397 is given as 0.18 in \citet{harris} and only varies between 0.185 and 0.191 across
the MUSE mosaic (see extinction map data from \citealt{Schlegel}, who combined results of IRAS and COBE/DIRBE).

The resolution of the available E(B-V) data is too low and would conceal possible structures on smaller scales
but could indicate a large scale gradient across the FOV which is not the case here.

\section{Extraction of absorption features in MUSE spectra}
\label{sec:extraction}
\subsection{Modeling of the stellar population and telluric correction}

Each of the stellar spectra were PSF-extracted simultaneously from the single data 
cubes. The process is described and successfully applied in Paper I.

In Paper I every extracted stellar spectrum of the globular cluster was fitted with a template spectrum based on PHOENIX 
models\footnote{The PHOENIX library used is described in \citet{Husser13}.}. The stellar templates
were scaled with a polynomial as linear fit to account for reddening and hypothetical
flux calibration inconsistencies in the commissioning data.
Furthermore, an individual telluric model was fitted to each spectrum, though 
we expect a rather homogeneous sky background for
each MUSE field of view.  Paper I describes this process in great detail.
The telluric absorption lines were fitted with models of varying \HHO\, and \Otwo/\Othree\, abundances \citep{tellurics}.
This work utilizes the 
residuals of the above mentioned template matching and sky fitting procedures.
Once the stellar templates and telluric absorption lines are subtracted, the only features 
that remain are hypothetical stellar lines not represented in the models, 
potential misfits of stellar templates, and
absorption features of the interstellar medium along the line of sight 
toward NGC\,6397. 
   \begin{figure}[h!]
   \centering
    \includegraphics[width=0.475\textwidth]{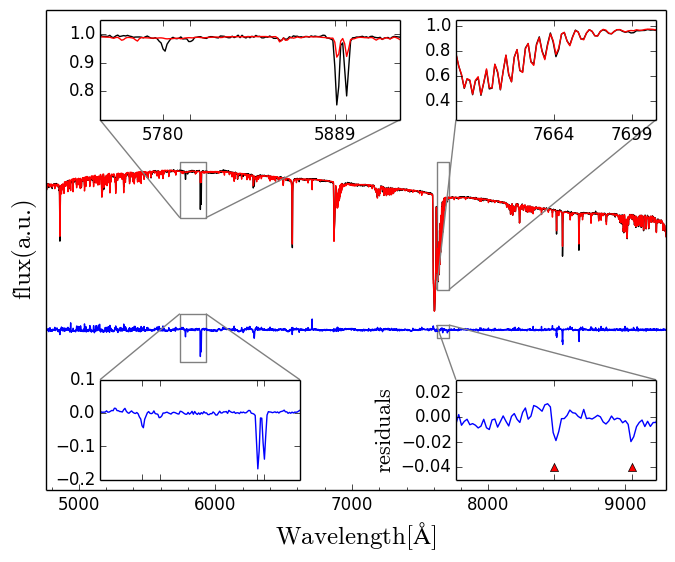}
    \caption{Steps of data processing. In {\it black} the data for one of the brightest stars
    in the field with a S/N $\sim$ 120. In {\it red} the best fit to the data including the telluric absorption line
    fit as well. In {\it blue} below the fitting residuals to scale with the plotted spectrum.
    The enlargement windows to the left, highlight the region around the 
  DIBs 5780,5797 and the \NaD doublet. 
    On the right hand side, the zoom illustrates the success of the telluric absorption line fit. We note that the
    zoom into the fitting residuals for this region is scaled by a factor of 10. For this example based on a single star
    of particularly high S/N, the \KI doublet stands out even for an individual stellar fit ({\it red} triangles).}
    \label{fig:reduction_steps}%
    \end{figure}
{ As a first iteration, a subset of the 19,000 spectra\footnote{The MUSE pointing mosaic
was arranged with some overlap, which is why the data set contains 18,932 spectra of
12,307 individual stars.} was selected based on a 
S/N threshold and a minimum stellar temperature.}

The temperature threshold was conservatively set to 4,000K.
{Colder stars begin to show strong molecular features and unresolved bands  
and it is practically impossible to get a reliable estimate of the local continuum.
For the final analysis, all stars with a S/N < 10 were excluded as their stellar template matching
lacks the precision to derive meaningful information from the residuals at the required level.}
That final sample contains 9,746 spectra.

{Figure \ref{fig:reduction_steps} illustrates the reduction process for an individual star.
The best template match is plotted on top of the observed data.
The method applied to fit the stellar templates does not
provide a stellar continuum model. From the perspective of the interstellar matter
along the line of sight, however, the stellar flux itself represents the continuum level and 
the residuals from the stellar fitting were then scaled with the final stellar template.
The resulting continuum spectrum is also shown.
After this step, the residuals range from 0, the former stellar flux, to -1, meaning 100\% of the stellar flux was absorbed.
This is illustrated in Fig. \ref{fig:reduction_steps} in the bottom left and right insets.
After adding an offset of one the processed residual spectrum has the same properties as a normalized absorption spectrum.}

\subsection{Improving S/N by tessellation}
\label{sec:voronoi}
For the general analysis, the individual residual-spectra, were co-added with error
weighting into Voronoi bins of equal S/N. The applied Voronoi algorithm is based on the procedure presented in
 \citet{cappellari03}.

Voronoi binning is a special kind of tessellation that solves the problem of preserving the maximum spatial resolution
of general two-dimensional data, given a constraint on the minimum S/N.
This automatically leads to somewhat homogeneous S/N per bin with varying bin-sizes. The uneven spacing of the 
stars as well as the strong variation in brightness of neighboring stars leads to varying shapes (but always convex)
of the individual bins.

As each spectrum was extracted from the 23 MUSE data cubes which were reduced to the same wavelength grid,
no re-sampling of the individual spectra was required for this step.

Our tessellation results in a bin size and thus a resolution which is coupled to the globular cluster via its star density.
When using a fixed grid across the FOV, we would couple the star count per bin to the underlying cluster and thereby
correlating the error per bin to the stellar distribution. As we deem it crucial to minimize any correlation
of the stars themselves with the observed ISM we chose the described tessellation.

We aimed for a S/N per bin of $\sim$ 150 which resulted in the 31 independent Voronoi bins.
Each bin contains $\sim$ 300 spectra. For a more detailed map  of \KI and DIB 5705 (see Sect. \ref{sec:highmap})
we realized a tessellation with 107 bins and $\sim$ 100 spectra per bin.
The chosen number of spatial bins is the result of a strict trade-off between S/N per bin
and spatial resolution.
 The spatial resolution of this approach for 
a globular cluster such as NGC\,6397 with many resolved stars is merely limited by the S/N of the stellar
spectra. The stars themselves are usually less than two arcseconds apart near the cluster core.
{ 
The 31 bins span about 20 arcseconds on average with equivalent width errors on the order of 8 m\AA\, and 
about 14 m\AA, for the maps with 107 bins and a S/N of $\sim$ 90 per bin.}
As these uncertainties in equivalent width are mostly a result of template residuals,
 the true uncertainty per measured individual equivalent width can vary slightly. 

\subsection{Measuring absorption features}
\label{sec:kiabs}
\begin{figure}[hb!]
   \centering
    \includegraphics[width=0.475\textwidth]{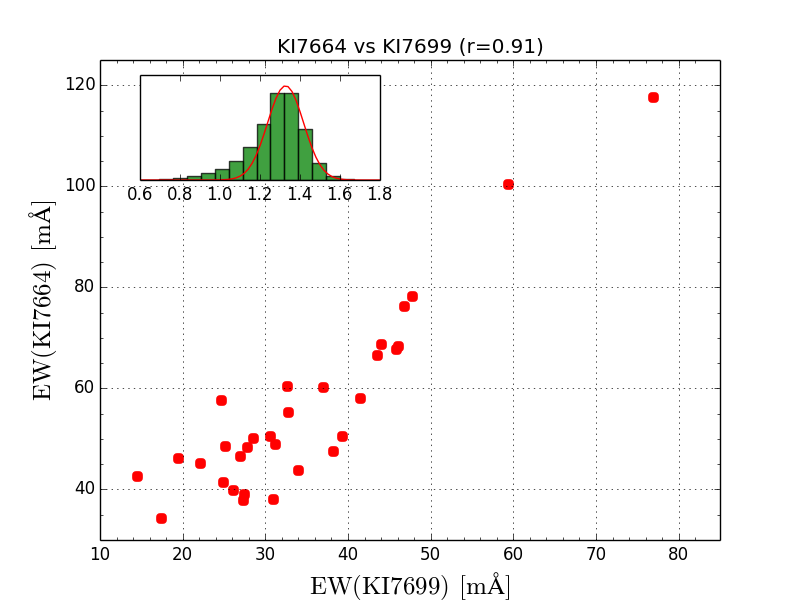}
    \caption{Equivalent widths per bin of \KI 7664 vs \KI 7699 for 31 bins.
    The correlation coefficient is 0.91. The inset in the upper left shows the distribution of the derived slope based on 5,000 bootstrap samples. 
        The Gaussian curve in red corresponds to 1.33 $\pm$ 0.09, though the ratio between the two lines is not constant
      which is also apparent from the asymmetric distribution.}
    \label{fig:kia_vs_kib}%
    \end{figure}

{ Several of the observable transitions of ISM species as well as DIBs fall into regions of strong sky contamination.
Successful modeling of the stellar contributions and the telluric absorption is key to precision measurements of the often quite weak (in the m\AA\, range)
equivalent widths (see, i.e., \citealt{Puspitarini13} and \citealt{monreal2017}).
While studies based on high resolution spectra can aim at modeling telluric absorption lines individually, these absorption bands
are not resolved by the MUSE spectrograph. Its great benefit is, however, to measure thousands of spectra simultaneously and
thus under the same conditions.
The accuracy of the telluric  absorption line fit is illustrated
in Fig. \ref{fig:kia_vs_kib}, in which we plot the equivalent width of both \KI components against
each other.
The line strengths for each doublet was fitted as a free parameter to use the ratio as sanity check 
as well as probe for density variations as mentioned in Sect. \ref{sec:maps}.
 \KI 7664 sits directly on a strong telluric absorption band, while 
\KI 7699 is hardly affected by this (see also Fig. \ref{fig:reduction_steps} and Fig. \ref{Fig:relativeo2}).
 At an equivalent width of $\approx$ 50 m\AA, both features are
extremely weak and seen at the MUSE resolution the absorption lines show a absorption depth of merely 1-2\% of the continuum level.
We are confident that
the fit of the sky model is sufficiently stable and accurate to recover the information on the stellar flux and
interstellar absorption in those regions of the spectrum. 
The inset in Fig. \ref{fig:kia_vs_kib} shows the distribution of the slope
for 5,000 bootstrap realizations of the dataset quantifying the correlation in a more robust way than the correlation coefficient 
of 0.91. Details on the fitting procedure for ISM features as well as DIBs are given in the Appendix.}

\section{Results on interstellar absorption}
\label{sec:results}

\subsection{MUSE integrated spectrum}
\label{sec:MUSE integrated spectrum}

After applying the procedure described in Sect. \ref{sec:extraction},
the former residual data behave like normalized flux 
spectra cleaned of stellar or telluric  absorption line features.

For the identification of ISM features, all stellar spectra were combined into an
error-weighted mean single high S/N spectrum.
The term signal-to-noise ratio is ambiguous in this context.
Interpreting noise as everything that does not contribute to our signal, basically residuals
after our ISM and DIB feature fit, we can derive a S/N
for the region 5600-5900\AA\, of $\approx$ 700 based on the standard deviation of the residuals.
That S/N also reflects systematic uncertainties in the template matching.
Stellar templates are not perfect and not all 
deviations are normally distributed and scale with $\sqrt{N} \sim 100$.
The photon noise itself, as a statistical quantity, is extremely low for our composite spectrum of almost 10,000 stars 
and the S/N is on the order of 2,000.

\begin{figure}[hb!]
   \centering
    \includegraphics[width=0.475\textwidth]{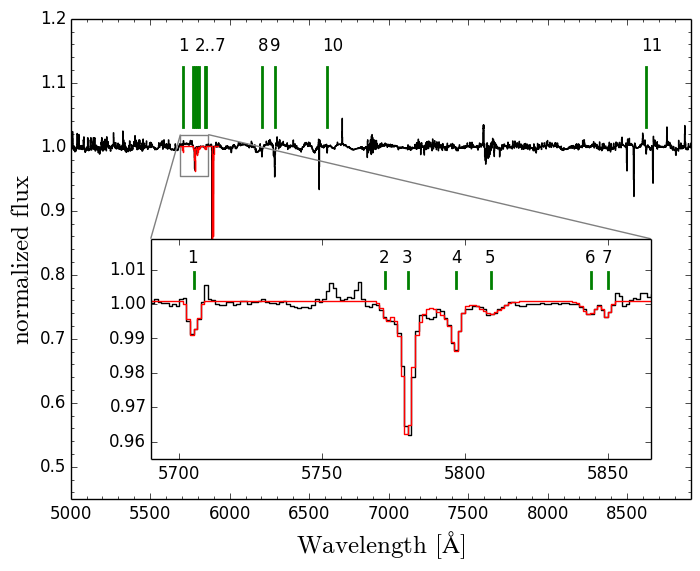}

    \caption{DIBs and \NaD in the combined residuals for all stellar fits. Namely, 
    DIBs 5705 (1), 5773 (2), 5780 (3), 5797 (4), 5809 (5), 5844 (6), 5850 (7),
    6203 (8), 6283 (9), 6613 (10), and 8621 (11). The DIBs are indicated
    with numbered {\it green} lines. The inset shows the DIBs 1-7. The {\it red} line is the multi-Gaussian fit for
    the selected region.}
    \label{fig:collapsed}%
    \end{figure}

The individual spectra were weighted by the square of their S/N.
An illustration with an enlargement of the wavelength of the resulting composite spectrum is shown in Fig. \ref{fig:collapsed}.
The S/N is $\approx$ 700 for the blue part and slightly lower for the red end beyond 7000\AA.
The equivalent width error due to the measurement uncertainties is
 $\sigma_{\mathrm{EW}}=\sigma_{\mathrm{flux}} \times \frac{\Delta\lambda}{N_{\mathrm{pix}}}$, where $\sigma_{\mathrm{flux}}$ is
the standard deviation for the flux, $N_{\mathrm{pix}}$ is the number of points covering the
absorption interval and $\Delta\lambda$ the corresponding wavelength range. For the MUSE sampling this
makes a factor of 1/1.25 = 0.8. The equivalent width error turns out to be about 2 m\AA\, for a DIB in the observed
 wavelength range.
Aside from the \NaD doublet and the two prominent \KI lines, numerous DIB 
features can be identified (see Table \ref{tab:ews}). Despite the moderate resolution of the MUSE spectrograph of $\sim$ 
2,200 - 3,500 in the range of observed DIBs, several DIB features are clearly resolved. 
{ The known large intrinsic width of the DIB features surpassing the LSF of MUSE was a motivation to carry out this study
with this instrument in the first place.}

All identified neutral gas species and DIB features in the composite residual spectrum were fitted as multiple Gaussians with 
the evolutionary algorithm described in \citet{quast05} and applied in \citet{wendt2012}. 
{ Their positions were fixed to avoid degeneracy with blended features.}
The equivalent widths were then derived from the modeled
Gaussians to account for possible overlap regions. 
The ISM features in the composite spectrum were fitted with Voigt profiles to derive the column densities.
The local continuum was fitted for each group by a low order polynomial.
{More details on the fitting procedure are provided in the Appendix.}
{For the two \KI lines as well as the \NaD doublet the radial velocity was fitted as a common parameter. 
The broadening parameter of the \KI doublet was tied for both transitions.
The same applies for the \NaD doublet.}

The equivalent widths of the identified DIBs, as well as of the \NaI and \KI lines are listed in Table \ref{tab:ews}.
For the combined spectrum the statistical uncertainty of the equivalent widths is on the order of 1-2 m\AA.
{ 
From the observed brightness temperature of the H\,{\sc i} 21cm emission
we estimated the total neutral gas column density $\log[N(\HI)]= 21.1$ and can compare it to the measured equivalent width of the 5780 DIB
for which a phenomenological relation is known.}

\begin{table}

    \caption[]{Measured equivalent widths on the combined spectrum. The statistical error is in the order of 1-2 m\AA.}
  \label{tab:ews}
    \centering
    \begin{tabular}{|l|r|r|}

\hline
    Line-ID     & EW [m\AA] & $\log(N)$ [$N$ in cm$^{-2}$]\\\hline
  DIB 5705      & 37 &\\
  DIB 5773      & 15 &\\
  DIB 5780      & 199 &\\
  DIB 5797      &  70 &\\
  DIB 5809      & 31 &\\
  DIB 5844      & 20 &\\
  DIB 5850      & 13 &\\
  DIB 6203      & 70 &\\
  DIB 6283      & 411 &\\
  DIB 6613      & 36 &\\
  DIB 8621      & 74 &\\\hline
  \KI 7664      & 60 &\\
  \KI 7699      & 38 &\\
  \KI           &    &11.3\\\hline
  \NaI 5889     & 525 &\\
  \NaI 5895     & 411 &\\
  \NaD          &     &12.5\\\hline
  \HI (main)           & & 21.1\\
  \HI (at -21\kms)       & & 19.9\\\hline
\hline
    \end{tabular}
      
   \end{table}

\citet{thorburn03} conclude, that DIB 5780 is  not likely to be saturated for equivalent widths
below 800m\AA, which is the case for our sight line. Then the galactic relationship (i.e., \citealt{Friedman11}) can be applied and it provides 
the estimate of $\log[\frac{N(\HI)}{\mathrm{cm}^{-2}}] = 19.05 + 0.92 \times \log[\frac{W5780}{\mathrm{m\AA}}]$ 
corresponding to $\log[N(\HI)]= 21.16$
 based on the DIB measurement alone, which is in very good agreement with our direct measurement of \HI.
Fitting the \HI emission with  multiple Gaussians allows an estimate of $\log[N(\HI)]=19.9$ for the
small component at $v_{\mathrm{helio}}\approx$ -21 \kms as also listed in the table.

\begin{figure}[hb!]
   \centering
    \includegraphics[width=0.45\textwidth]{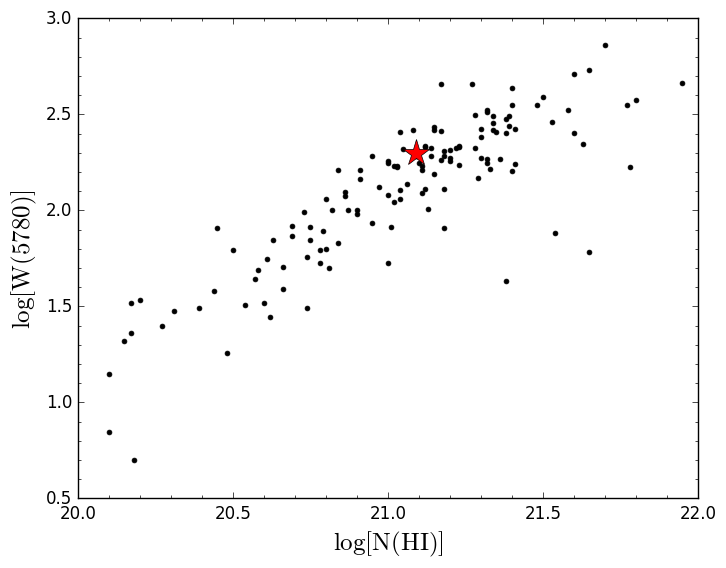}
    \caption{W(5780) vs. N(\HI).
    The data for the composite spectrum as listed in Table \ref{tab:ews} (red star)
    overlaid to corresponding data for the Milky Way as published in \citet{welty06}, \citet{Friedman11} 
    and   \citet{welty01} (as scatter plot in black).
    Measurement uncertainties of those data vary between 3m\AA\, and 20m\AA\, for DIB 5780 and
    are omitted here for clarity. For our measurement the statistical error is 
    about 1-2m\AA\, and lies within a fraction of the star symbol.}
    \label{fig:welty}%
    \end{figure}

Figure \ref{fig:welty} relates these measurements to those by \citet{welty06}, \citet{welty01}  and
\citet{Friedman11} in the Milky Way. The MUSE data allow for excellent measurements of the DIB equivalent widths.
The comparison to other DIB data in Fig. \ref{fig:friedman} further verifies the method
utilized here to measure the equivalent widths in composite fitting residuals of cluster stars as the 
derived absolute equivalent widths for different DIBs lie within the expected correlation to the strong DIB 5780.

%
\subsection{Spatially resolved \NaI, \KI, and DIBs}

\subsubsection{Maps}
\label{sec:maps}
\begin{figure*}[hbt!]
   \begin{center}
\includegraphics[width=0.7\textwidth]{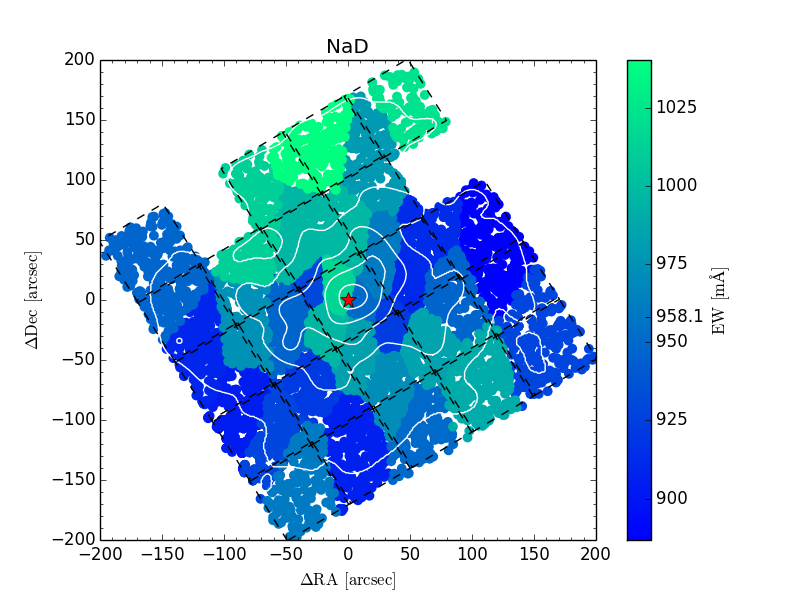}\\
\includegraphics[width=0.7\textwidth]{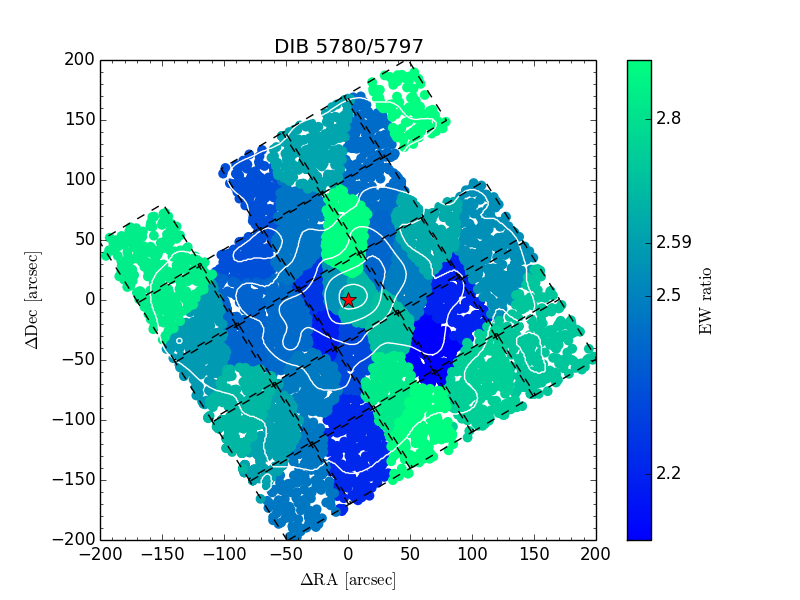}
 \caption{Equivalent width map for the \NaD doublet ({\it top}) and the DIB ratio 5780/5797 ({\it bottom}).   
 All selected cluster stars are colored in 31 groups according to their Voronoi-bin.
The color reflects the measured total equivalent width of the \NaD doublet for each composite spectrum 
per bin with an average of about 960 m\AA.
The center of the globular cluster is at coordinate origin and marked with a red star.
Brightness contours of the globular cluster are shown as white lines. For the contour plot, the white light image was
smoothed to a resolution of 20 arcsec.}
\label{Fig:bigmap}%
\label{Fig:dibratiomap}
\end{center}
    \end{figure*}

\begin{figure*}[hbt!]
   \begin{center}
  \includegraphics[width=0.45\textwidth]{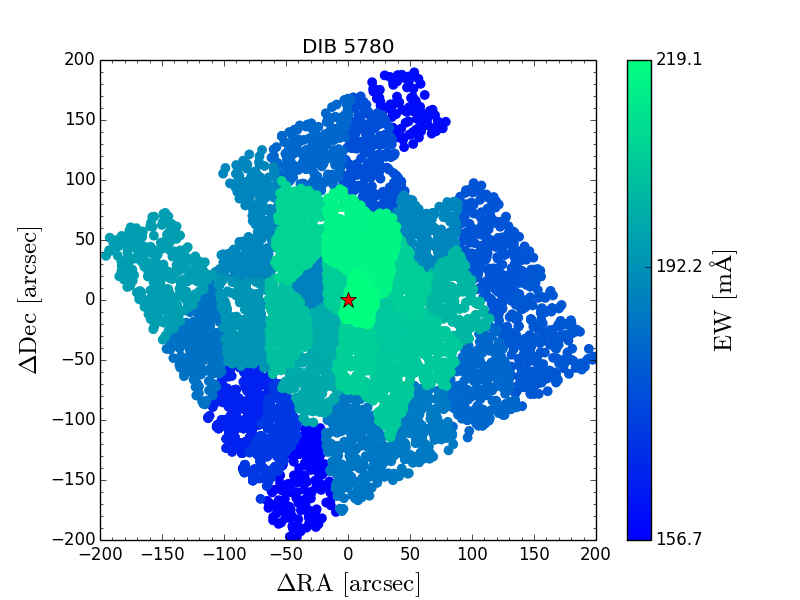}
\includegraphics[width=0.45\textwidth]{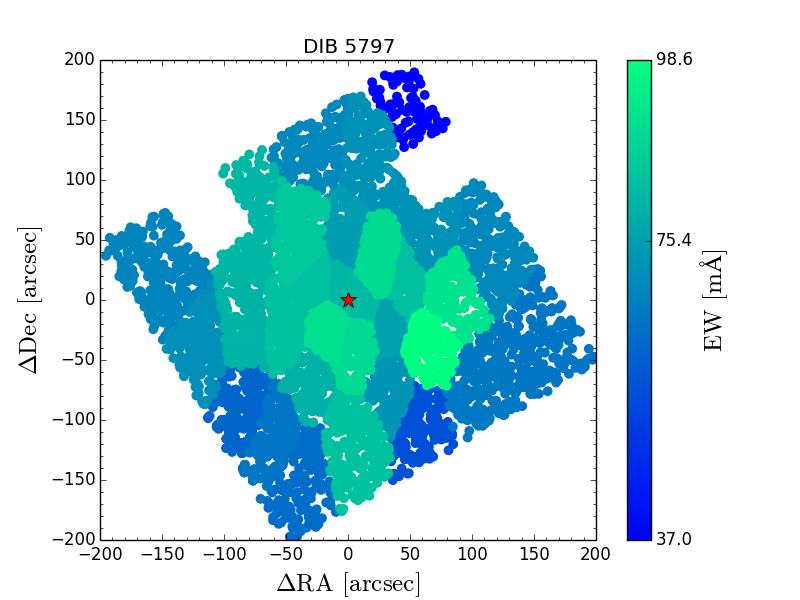}\\
\includegraphics[width=0.45\textwidth]{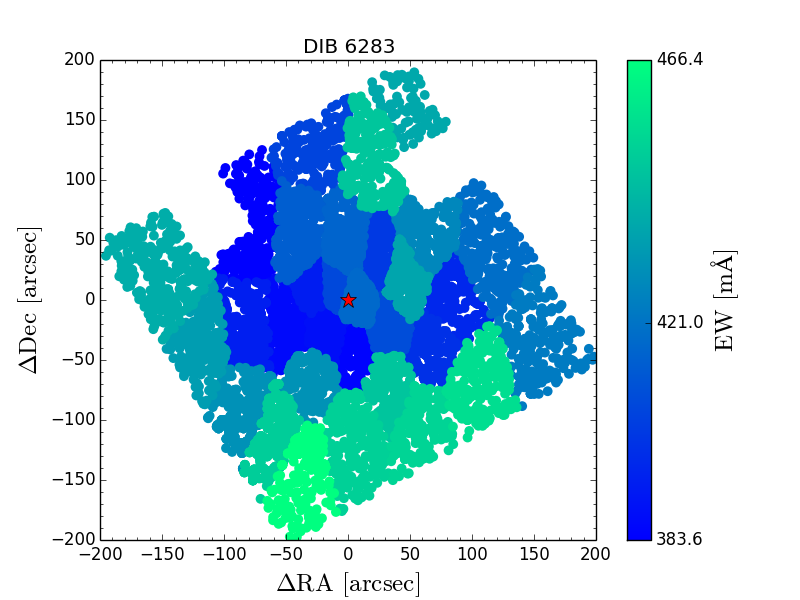}
\includegraphics[width=0.45\textwidth]{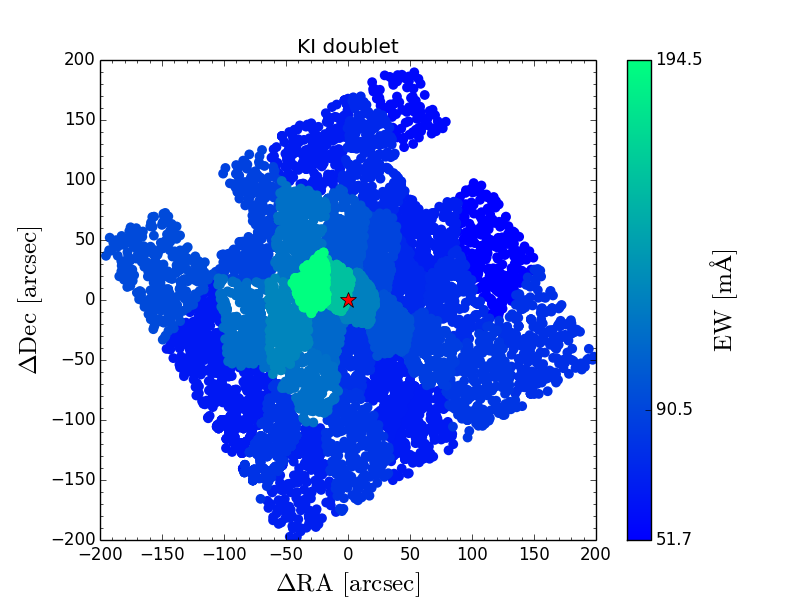}\\
\includegraphics[width=0.45\textwidth]{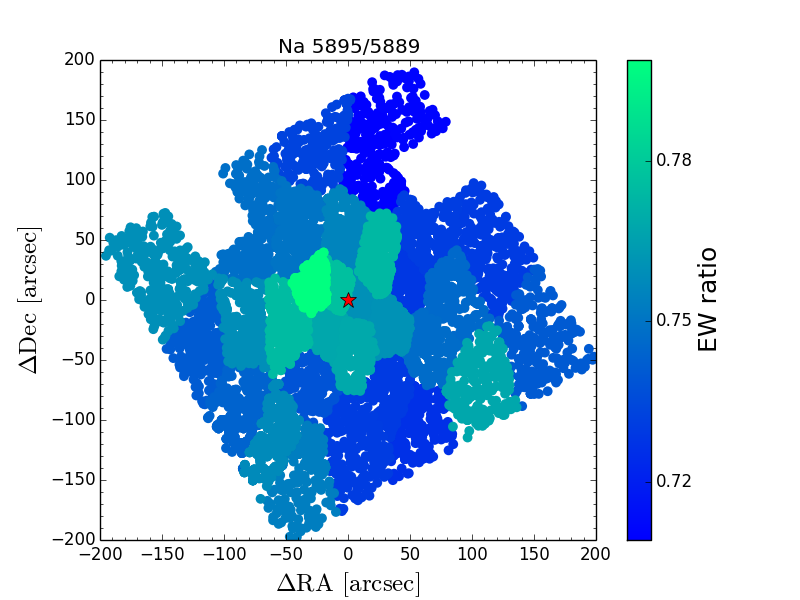}
\includegraphics[width=0.45\textwidth]{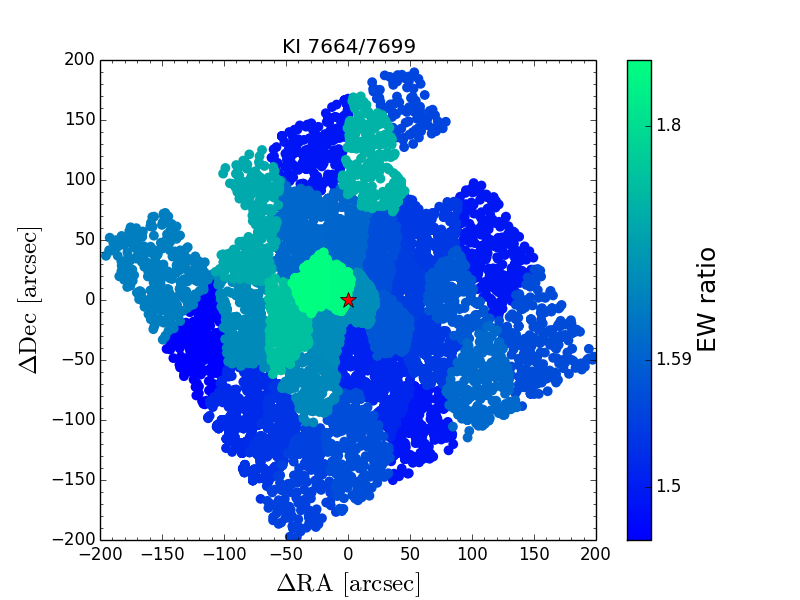}\\
    \caption{{\it Top:} Equivalent width maps with DIBs 5780, 5797;
  {\it Center:} DIB 6283 and \KI.
 {\it Bottom:} maps of doublet line ratios for \NaD and \KI, indicating optical depth.
 For each map, the $\approx$10,000 selected spectra 
   were combined into 31 Voronoi bins. The color bar shows the equivalent width range as well as the average value
  for the species. The uncertainty is in the order of 8 m\AA.}
              \label{Fig:maps}%
\end{center}
    \end{figure*}

In the previous section we saw that the measurements on the composite spectrum of almost 10,000 cluster stars 
are in very good agreement with other measurements of individual sight lines in the Milky Way.
Rather than finding universal correlations between species, we are in the position to be able to  probe correlations
between DIBs on very small scales and thus likely in a physically connected absorber.
{ Figure \ref{Fig:bigmap} shows a spatial map of \NaD and of the DIB ratio 5780/5797 (this ratio
is thought to be sensitive to the UV radiation field: \citealt{Ensor2017}).}
The implications are discussed in Sect \ref{sec:dibsandism}.
 The measured total equivalent width of \NaD shows no correlation with the
brightness of the GC which would otherwise indicate some underlying systematics in the analysis.
{ 
Further spatial maps are shown in Fig. \ref{Fig:maps}. The large spatial variations up to a factor of four (\KI) or two (DIB 5797) stand out.}
DIB 6283 shows a distinct anti-correlation 
 to most of the other observed species. Where DIBs 5780, 5797 and \NaD are strongest,
the strength of DIB 6283 drops. This observation is discussed in Sect. \ref{sec:anti}.
{
While the composite equivalent width correlates to the total column density of the gas, the chosen line ratios trace
the optical thickness of the gas and it shows well that the abundance of neutral potassium correlates well with 
that of \NaI.
}
The equivalent width of the \KI absorption shows a rather clumpy distribution in Fig. \ref{Fig:maps} and we
prepared maps at higher resolution (at the cost of a drop in S/N) of those absorption features in Fig. \ref{Fig:highres_ism}.
It reveals a \KI region with a size of $40''\times40''$.
Figure \ref{Fig:highres_ism} demonstrates that the results in this paper are stable and
independent on the specific realization of Voronoi-binning. As expected, the strengths of the absorptions features peaks
at higher values for the high resolution, resulting in a slightly higher average equivalent width per bin as well.
The given errors represent the statistical variance rather than the uncertainty with respect to the true, potentially 
unresolved peak of the absorption.

DIB 5705 appears to have a relatively sharp edge and shows a slightly softer gradient toward the center of the map.
 {The visible structure is not aligned with the individual MUSE pointings on the sky (see Fig. \ref{Fig:o2} in the Appendix),
 which would indicate, for example,
 sky removal problems or other systematics that scale with exposure time or depend on environmental conditions.}
 { DIB 5705 and \KI show the strongest and sharpest variations in equivalent widths within the field
of view, although the maximum is not located at the same position for these two tracers.}
The \NaD strengths show a filament-like structure and a general gradient which does not coincide spatially with
sky background or cluster brightness at all.
We consider this further confirmation that the separation of stellar content, telluric  absorption lines and interstellar medium was
very effective.
All visual structures consist of several bins and thus are not dominated by putative peculiar individual stars.
{These initial results encourage us to perform a more in-depth study based on a larger sample of GC once available.}
\subsubsection{Comparison between \NaI and \KI absorption}
\label{sec:compareNI}
 For the 31 bins in Fig. \ref{fig:kia_vs_kib},
 300 stars (on average) were co-added per bin prior to the EW measurement of the \KI doublet lines.

\begin{figure}[h!]
   \centering
   \includegraphics[width=0.475\textwidth]{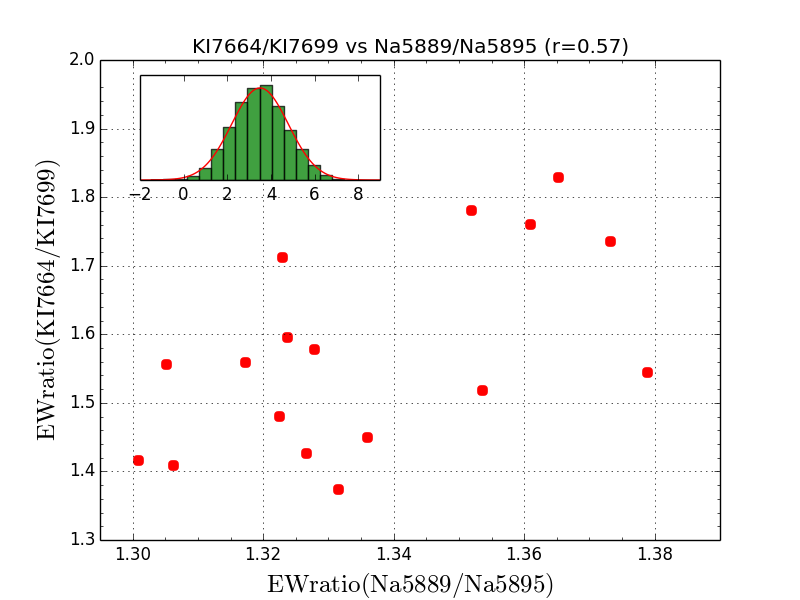}
    \caption{Ratio of \KI components vs \NaD components.
      The inset in the upper-left shows the distribution of the derived slope based on 5,000 bootstrap samples. 
        The Gaussian fit in red corresponds to 3.5 $\pm$ 1.3.}
     \label{fig:ki_vs_na}
\end{figure}

These two \KI features were fitted as independent spectral lines to use them as a qualifier of the telluric correction and
to correlate the ratio with other quantities. 
The \KI 7664 vs \KI 7699 ratio is approximately two for optically thin gas, and approaches approximately one for
optically thick gas. We observe a range between 1.4 and 1.8.
Following \cite{Williams1994}, the optical depth $\tau$ of the 7664 \AA\, line  can be calculated from 
\begin{equation}
\frac{I_\mathrm{7664}}{I_\mathrm{7699}}=\frac{1-e^{-\tau}}{1-e^{-\tau/2}}
\end{equation}
and considering the lowest equivalent width ratio measured for the \KI lines to represent their
intensities $I_\mathrm{7664,7699}$, we derived a value for the optical depth of $\tau \approx 1.8$
 for the densest region and a $\tau$ as low as $\approx 0.4$ for the surrounding environment.
For the \NaD ratio a similar rule applies. For optically thin gas, the ratio of the \NaD doublet 5889/5895 would equal 2.04. 
The effect on the equivalent widths of \NaD is much lower as it is mostly saturated already. 
{ The measured line ratio ranges between 1.4 and 1.3 (see Fig. \ref{fig:ki_vs_na}) and reflect the opacity of the gas.
Under the assumption of comparable absorber sizes and a uniform broadening parameter the opacity is proportional to the gas density. 
}
Both independent indicators for gas density correlate in our data set as indicated in Fig. \ref{fig:ki_vs_na}.
They are separated by $\sim$ 1800 \AA\, and are affected by different atmospheric species.
The inset in the upper left shows the distribution of the derived slope based on 5,000 bootstrap samples. 
        The Gaussian fit in red corresponds to 3.5 $\pm$ 1.3.
{The Spearman's rank correlation is 0.45.
A two-sided  hypothesis test whose null hypothesis is that two sets of data are uncorrelated yields a probability of only 0.06
and this scenario thus can be rejected at the 2-$\sigma$ level.}

The \KI doublet probes gas with nearly the same ionization state as \NaI, 
but is more likely to be optically thin (the abundance is a factor of about 15 lower than that of \NaI,
while the two doublets have similar oscillator strengths).
The ionization potentials of \NaI and \KI are only 5.1 eV and 4.3 eV, respectively.
These species should primarily probe the \HI\, and \Hmol ISM phases or represent otherwise non-dominant ionization states
in the more diffuse warm neutral gas.
Since \KI and \NaI have very similar ionization potentials, and show a similar 
grain depletion pattern \citep{savage1996}, the expected ratio of the \NaI and \KI column densities
 should be $\approx$ 15 for gas with a solar Na/K ratio.
For the collapsed spectrum we determine column densities of $\log[N(\NaI)]=12.46$ and $\log[N(\KI)]=11.3$, with a 
\NaI/\KI ratio of 14.5, which is in excellent agreement with the prediction.
{In Sect. \ref{sec:uvesmuse} we discuss the accuracy to which this ratio can be derived from low resolution MUSE data.}
\subsection{Comparison of DIBs, \NaI, and \KI}
\label{sec:dibsandism}
\begin{figure}[h!]
   \centering
   \includegraphics[width=0.475\textwidth]{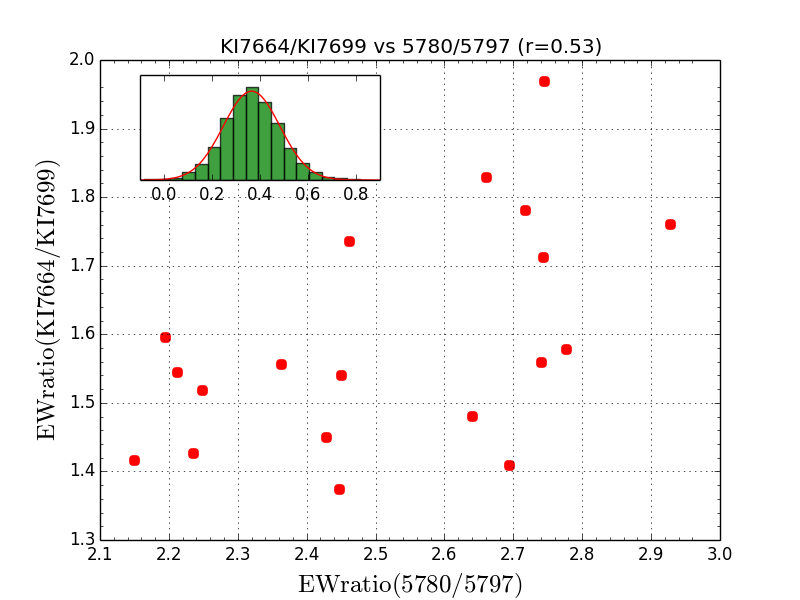}
    \caption{Ratio of \KI components vs DIB ratio 5780/5797, i.e., ionized/neutral gas.
        The inset in the upper-left shows the distribution of the derived slope based on 5,000 bootstrap samples. 
        The Gaussian fit in red corresponds to 0.37 $\pm$ 0.12.}
        \label{fig:ki_vs_dib}
   \end{figure}

We find a correlation of \KI 7664/7699 to the DIB 5780/5797 ratio as illustrated in Fig.
\ref{fig:ki_vs_dib}. The correlations in Figs. \ref{fig:ki_vs_na} and \ref{fig:ki_vs_dib} are significant according to a Student $t$ test
 with $t > t _{\mathrm{critical}}$, where $t=r\sqrt{(N-2)/(1-r^2)}$, with
correlation coefficient $r$ and $N$ data points (with in this case $N-2$ degrees of freedom).
A bootstrap analysis based on 5,000 samples yields a slope of 0.37 $\pm$ 0.12.
{The Spearman's rank correlation is 0.54.
A two-sided  hypothesis test whose null hypothesis is that two sets of data are uncorrelated yields a probability of only 0.016
and this scenario thus can be safely rejected.}

According to, for example, \citet{a7} and \citet{vos}, the carrier(s) of DIB 5780 do
 not change ionization easily (it is either already ionized,
or neutral and hard to ionize), while DIB 5797 is thought to be caused by a neutral species, easily ionized. Thus, a low
DIB 5780/5797 indicates a denser, more shielded region that is less exposed to UV radiation.
{\citet{Ensor2017} also identify this ratio as a good approximation for the level of UV exposure in their
analysis and they find that DIB 5797 itself correlates very well with the overall strengths of DIBs.}
The correlation of the DIB 5780/5797  ratio with \NaI opacity is expected but was not yet
 observed within a single absorbing region. The spatial mapping of the DIB ratio in the bottom panel of Fig. \ref{Fig:dibratiomap} appears
to trace a filamentary structure across the field of view. The lowest value of the ratio is in the region that bears
the strongest \KI absorption and highest optical depth. The highest ratios are seen just slightly shifted to the
right ($\approx 30$ arcsec) of the filament like structure seen in the \NaD equivalent maps in Fig. \ref{Fig:bigmap}.
This could indicate some ionizing radiation dominantly from one side (bottom right in the coordinate system used),
causing the observed DIB 5780/5797
ratio while the dense regions leftward of this filament are still shielded and less affected.

\section{Discussion}
\label{sec:discussion}

We have no direct information on foreground stars, in particular not on individual distances. In Paper I and II,
cluster members were identified based on a number of selection criteria. The sample used in this analysis does not consist
exclusively of member stars. A comparison with a composite spectrum of likely non-member stars, yielded no significant
difference in EW strengths to a members-only selection. This further supports the assumption of the absorber being relatively close.
We expect more conclusive distance estimates for some of our other targets.
For other globular clusters in our Milky Way, in particular those near the Galactic bulge, the distinction between foreground stars and 
cluster members will be crucial. Distance measurements by GAIA will contribute greatly to a better understanding of the involved
distances for those targets.
{
A more comprehensive interpretation of the new findings will be discussed
on the basis of a considerably larger data sample in a follow-up paper that will utilize up to 25 globular cluster targets
  observed as part of the MUSE guaranteed time observations (PI: S. Dreizler). }

\subsection{UVES vs MUSE}
\label{sec:uvesmuse}
{ A conclusive direct comparison of the MUSE data with the single spectrum taken with UVES is not possible since the line of sight toward the 
individual star lies outside of the mosaic observed with MUSE.}

In the single VLT/UVES exposure of NGC\,6397-T183 
we measure equivalent widths of 85 m\AA\, for DIB 5780, and 200 m\AA\, for DIB 6283,
both modeled with Gaussian functions. These values are considerably lower than for the combined spectrum
toward the cluster core that are listed in Table \ref{tab:ews} but the line ratio measured in the UVES data agrees very well with the expectation
(see Fig. \ref{fig:friedman}, bottom right).

For the UVES spectrum we derived a total column density of $\log[N(\NaI)]=12.6$ based on Voigt profile
fits of the three components at $v_{\mathrm{helio}}\approx$ -26 \kms, -10 \kms and at rest (see Fig. \ref{fig:uves1}).
This matches the prediction given in \cite{Ellison2008} based on the equivalent width of DIB 5780 in galactic sources as well:
$\log[\frac{N(\NaI)}{\mathrm{cm}^{-2}}]=10.51+1.1\times\log[\frac{W5780}{\mathrm{m\AA}}]=12.63$.

However, this relation estimates a total column density of \NaI\, for the MUSE FOV based on the equivalent width for the combined spectrum
of 199 m\AA\, for DIB 5780 of $\log[N(\NaI)]=13.04$.
The fit to the MUSE data yielded $\log[N(\NaI)]=12.5$, though.
While we can measure equivalent widths quite well, we can not expect to obtain accurate results from Voigt profile fits of unresolved
absorption features in particular when they are not optically thin, which is indicated by
the observed line ratios of the \NaD\, doublet.
For the combined MUSE data we obtain a ratio of 411m\AA\ / 525m\AA\, = 0.783.
The UVES data yields 386m\AA\,/ 511m\AA\, = 0.755 indicating a slightly lower optical depth.

The \NaI/\KI ratio of 14.5 mentioned in Sect. \ref{sec:compareNI} thus is critical as
fits by Voigt profiles of \KI\, and \NaI\, lines observed at this low resolution 
are not expected to provide accurate values for column densities. 
{ 
\NaI lines are not optically thin and therefore, $N(\NaI)$ is expected to be underestimated; 
 to a lesser extent, this is true also for $N(\KI)$.}
\subsection{Location and size of the absorbing cloud}
\label{sec:location}
The high-resolution UVES spectra as well as the radio data for \HI\, indicate that we observe one dominant absorption component
toward NGC\,6397. The component blueward of the main feature is more than an order of magnitude weaker in \HI.

Our Sun and nearby stars reside in a region of ionized
material referred to as the Local Bubble (\citealt{Frisch2011} and references therein). 
 The
edge of the Local Bubble can be traced by the onset of \NaI and \CaII
absorption, indicators of colder material. 
According to  \citet{Sfeir1999} or \citet{Welsh2010} this edge begins
anywhere from 65 to  250 pc depending on the observed direction.
Within the Local Bubble, isolated clouds of warm, partially
ionized gas are observed \citep{Redfield2008, Malamut2014}.

Assuming a distance of about 100 pc for the absorber, which would put it roughly at the rim of the Local Bubble would
result in a projected size of our MUSE mosaic of about $30,000 \times 30,000$ AU.
The compact, presumably cold \KI core as illustrated in Fig. \ref{Fig:highres_ism} would
translate into a region of merely 4,000 AU in projected diameter.

Even though works such as \citet{bailey15} suggest compact self-shielded cloudlets to be present close to the sun 
as well, absorbers such as the observed one with matching \HI\, detection are unlikely to originate  within the Local Bubble.

{A comparison with the 3D ISM map in \citet{Lallement} shows compact regions of dense ISM at a distance of $\sim$ 100 pc
roughly in the direction of 
the observed globular cluster.
More detailed maps in \citet{Puspitarini14} reveal that the line of sight toward NGC\,6397 passes near a cluster of patchy ISM blobs.
Bringing together such complementary  studies will be much more beneficial for the complete sample of 25 globular clusters.}
{ At the present spatial resolution, these 3D maps support our interpretation but are insufficient to pinpoint   
gas clouds toward NGC\,6397.}

\subsection{Small-scale structure}
{
In Sect. \ref{sec:maps} we demonstrate that the approach of using the residuals of template matching 
$\sim$ 10,000 stars is successful in reconstructing the ISM absorption and preserving the spatial information
in Voronoi bins. 
Former works generated maps of ISM structures at parsec scales, while variations detected on smaller scales usually rely on 
few individual stars. 

Figure \ref{Fig:dibratiomap} shows filamentary structures that extend over the whole mosaic of $300''$ diameter, while 
the thinnest parts are no broader than $\approx 20''$. Following the argumentation in the former section, these translate to
coherent filamentary structures of $\approx 2,000$ AU in width stretching out over the full 
covered range of $\approx 30,000$ AU. The obtained MUSE data are unique in that respect. We provide a full 2D map at
a spatial scale that appears to be of the order of magnitude to trace these strong variations. Large-scale variations are
 well known and certainly exist in the scales that are observed already in \HI.

For example \citet{Bates1991} describe a rapid increase in gas-column density toward M\,22 on projected scales of 100 pc.
{ 
On very small scales, variations in the general ISM on subparsec scales have been found in previous
 studies of wide binaries. \citet{Watson96} report changes of a few percent at scales of some hundred AU, whereas
 \citet{Meyer96} observe stronger variations up to a factor of three at distances of about 6600 AU, which is comparable to our findings. 
 }

{ \citet{Andrews2001} observe variations of factor between four and seven in the column densities of \NaI in spectra of cluster stars with the DensePak fiber-optic array
and interpret them as small-scale turbulent variations.
They reach a spatial resolution of about $4''$, corresponding to a few thousand AU at the distance of the absorber which is comparable with the work in this paper.
However, they lack the large FOV to generate consistent maps on these scales. }

All in all there is convincing evidence of structures in the ISM at a large range of scales.
Filament structures appear to be accessible with our approach and we intend to apply this method
to more than 20 GCs to identify universal scale sizes and very local trends of \NaI, \KI, and DIBs
to derive properties of the typical environment for the individual species.
}
\subsection{DIB correlations}
Correlations between individual DIB strengths have
been used to identify features that respond similarly to
differences in the diffuse cloud environment. 
Identifying such correlating subsets of DIBs provides a valuable
spectroscopic constraint on the identity of the carrier(s).
\citet{adam2005} point out that any group of features arising from a particular carrier, or set of chemically related carriers, must maintain
the same relative intensities in all lines of sight. 
Directly comparing DIB equivalent width ratios within the same observations avoids the potentially complicating factor of normalizing individual DIB equivalent
widths which is a crucial source of systematics when comparing data of different sightlines.
   \begin{figure}[h!]
   \centering
   \includegraphics[width=0.475\textwidth]{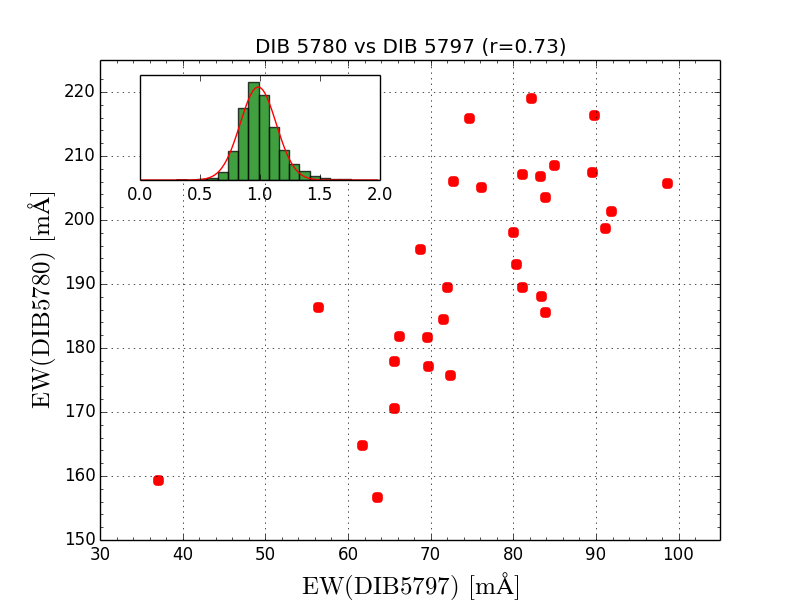}
    \caption{Equivalent widths of DIB 5780, vs 5797.
The inset in the upper-left shows the distribution of the derived slope based on 5,000 bootstrap samples. 
        The Gaussian fit in red corresponds to 0.99 $\pm$ 0.15}
       \label{Fig:5780_5797}
   \end{figure}
Fig. \ref{fig:friedman} shows the derived equivalent widths in the context of the universal DIB correlations detected
in many sightlines. While we are in excellent agreement with the general trend, we observe such correlations in our comparably
small field of view as well. The common correlation of DIB5780 to DIB 5797 is plotted in Fig. \ref{Fig:5780_5797}
for 31 bins within our FOV. The correlation coefficient is $r=0.73$ and a linear fit to 5,000 bootstrap-realizations
of the sample show a slope of $0.99\pm0.15$. The tight correlation of DIB 5780 and DIB 5797 appears to be
universal on all observed scales.
{
Former measurements of the DIB 5780 to DIB 5797 relation as summarized in \cite{Friedman11} yield
a very consistent ratio of $\approx$ 2.3:1 with an offset of about 30 m\AA. While the measurement on the combined data
reproduces that relationship exactly, the observed trend of our 31 spatial measurements within the physical absorber
shows a slope of almost exactly one. Figure \ref{Fig:fitfried} shows the two data sets and their best linear fit.
For this single pointing it is hard to judge whether this provides a deeper insight into the nature of the corresponding
chemically related carriers.
That is one result of this pilot study that we hope to explore further in a much larger 
investigation based on up to 25 globular clusters. It will show whether the local trend in fact deviates from the universal mixing.
}
 Interpreting the DIB ratio 5780/5797 as proxy for gas density (see Sect. \ref{sec:dibsandism}) 
it can be derived that the regions of higher optical depth also trace the denser gas regions.
This can also be seen in Fig. \ref{fig:ki_vs_na} and Fig. \ref{fig:ki_vs_dib}.

   \begin{figure}[h!]
   \centering
   \includegraphics[width=0.5\textwidth]{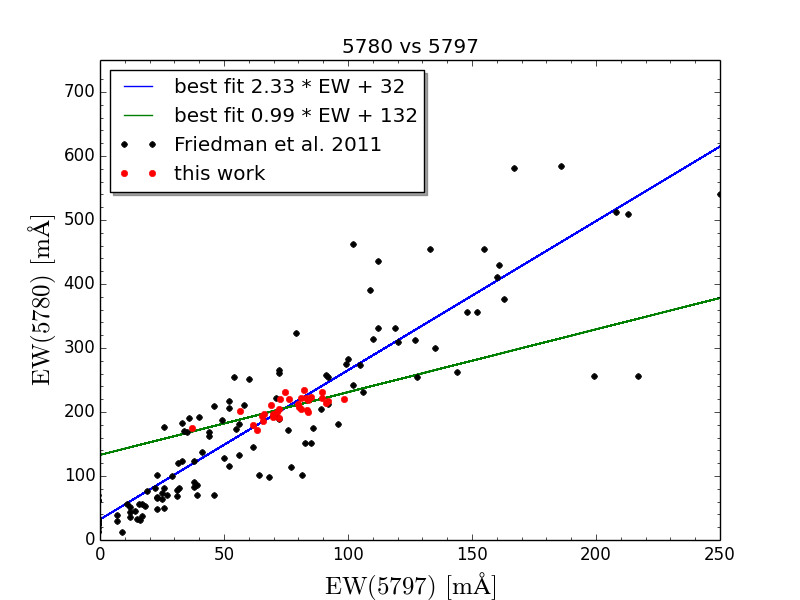}
      \caption{ Equivalent widths of DIB 5780, vs 5797 from \citet{Friedman11} ({\it black}) and this work ({\it red})
      along with a linear best fit. While the average ratio of DIB 5780 to 5797 matches the universal trend perfectly, the observed
      relation within the absorber differs (see Fig. \ref{Fig:5780_5797}).}
         \label{Fig:fitfried}
   \end{figure}
{
The filament structure of Fig. \ref{Fig:bigmap} is not seen in \KI which in principle should be more sensitive to structure.
The bottom map in Fig. \ref{Fig:maps} as well as that in Fig. \ref{fig:ki_vs_na} demonstrate that the doublet ratios, 
which indicate optical depth, trace each other rather well. The total abundances measured as equivalent widths, however, do not show
the same structures. Reliable measurements of the column densities are not feasible at the given spectral resolution. \citet{Graham2015} observe a behavior of interstellar \KI vs \NaI that could be explained by it being more sensitive
to UV radiation than \NaI. \cite{a7} describe in their study of small-scale structure in the interstellar medium that
\NaI and \CaII were present throughout the cloud, but \CaI was only present in the denser regions with \KI being intermediate
between \NaI and \CaI. Thus, one possible explanation for our observations is that the region of stronger \KI absorption
traces an area of a certain density and shielding from UV photons, while the broader area is less
favorable for \KI than \NaI. The structures of \NaI and \KI will be an important
part for future analyses. At present we cannot explain this behavior satisfactorily.
The coming follow up study of numerous GC observations of MUSE will unfold if such a local trend can be observed commonly.}

\subsection{Anti-correlation between DIB 6283 and other DIBs}
\label{sec:anti}
Unexpectedly, all mentioned DIBs anti-correlate with DIB 6283.
{ Other studies observe a universal correlation between DIB 6283 and DIB 5780 that is similar in strength to that of
DIB 5797 to DIB 5780  (e.g., \citealt{Friedman11}).}
This section will discuss the meaning and credibility of such an anti-correlation.

The maps and ratios for \KI and \NaI as well as for the other DIBs verify the success
of the handling of the telluric absorption lines.
The  templates used for the telluric absorption lines, are based on the best match to a model
grid with only 
the abundance of \HHO\, and \Otwo/\Othree\, as free parameters. 

While telluric \Otwo\, overlaps with DIB 6283 and the \KI 7664, 7699 features,
 the fitted \Otwo\, abundance is not able to induce an
 anti-correlation between DIB 6283 and \KI. 
 Furthermore, the correction for \Otwo\, works remarkably
 well, as demonstrated by the tight correlation in Fig.
\ref{fig:kia_vs_kib}, which gives us confidence in the result for DIB 6283 as well.
Figure \ref{Fig:o2} shows the strengths of  \Otwo\, in the best fit sky model.
The inner pointings that show stronger \Otwo\ absorption, belong to five consecutive
exposures during one night as illustrated on the bottom panel of Fig. \ref{Fig:o2}.
{ This impacts some of the central fields of the map, which also showed the weakest DIB 6283 strengths.}

\begin{figure}[h!]
   \centering
\includegraphics[width=0.45\textwidth]{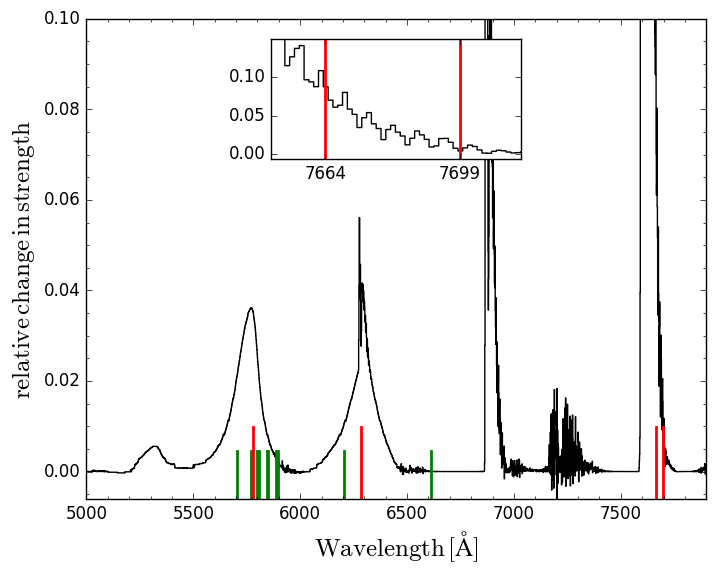}
      \caption{Maximum difference in modeled \Otwo\, absorption relative to the total mean of all exposures.
Positions of observed DIBs indicated in {\it green}. The {\it red} lines mark DIB 5780, DIB 6283 and the \KI\, doublet at 7664, 7699\AA,
which is also shown in the inset at the top. }
         \label{Fig:relativeo2}
\end{figure}

Figure \ref{Fig:relativeo2} shows the difference of the two extreme cases in the data set: two exposures for which 
the sky model contained the strongest and the weakest \Otwo\, while other telluric model parameters remained similar.
The modeled absorption for \Otwo\, varies within a few percent only and affects DIB5780, 5797 and DIB 6283
in an almost identical manner, whereas the effect for \KI 7664 is about twice as strong and less than half as strong for \KI 7699. 
{ The strength of the telluric feature blending with DIB 6283 is of the same order of magnitude but a little weaker than the DIB itself for this particular target with considerably high extinction. A variation of a few percent here as indicated in Fig. \ref{Fig:relativeo2} alone can not explain the measured range of DIB 6283 from 380 m\AA\, to 470 m\AA. For comparison: the telluric features in the region of the \KI\, lines are about 30 times stronger than the \KI\, features themselves which show no unusual behavior (see section \ref{sec:kiabs}).}
At the MUSE resolution, DIB 5705 is strongly
blended with a stellar line as well but its equivalent width 
shows no correlation with the background sources at all (see Fig. \ref{Fig:highres_ism}), despite being on the
 order of the stellar feature and we therefore have confidence that our measurement is reliable.

The stellar \NaD lines are excluded from the stellar fit and simply scaled
with the model.
The equivalent width of DIB 6283 in the composite spectrum matches the expectations exactly which rules out
notable systematic offsets or misfits.
{ It is possible that  the universal behavior of DIB 6283 is
just governed by the amount of interstellar material 
and  mixing of chemical abundances in general,
whereas we are tracing the DIBs relative to others within a physical environment that shows inhomogeneities in density, temperature and shielding of UV photons on small scales.}
DIB 6283 could favor regions of ionized gas (see also \citealt{Ensor2017}). 
\citet{Milisavljevic14} noted diametrically opposed behavior of DIB 5780
and DIB 6283 over time in the spectra 
toward a fading super-nova.
In the Milky Way, stronger UV radiation fields tend to lead to weakened DIB 5797 but enhanced DIB 5780
 and DIB 6283 (e.g. \citealt{Ehrenfreund1995}).

Assuming stronger UV radiation from the bottom-right in the FOV could explain why
we observe higher equivalent widths for DIB 6283 toward that direction where it gets ionized first.
{ The same applies for the DIB 5780/5797 ratio, which is strongest toward that direction
(see Fig. \ref{Fig:bigmap}, {bottom} and Fig. \ref{Fig:maps}, {left center}).}

The amplitude of the variation in the field of view is $\approx 80$ m\AA\, for DIB 6283.
Inaccurate modeling of stellar lines in this particular spectral region is still a possible source for systematics though
the distribution of measured equivalent width strengths is not directly linked to any quantity of the cluster itself.
Future analyses will help to confirm or refute this result. 

{
The map of DIB 5705 in Fig. \ref{Fig:highres_ism} also shows no clear correlation
with the other species. While it is affected by a stellar line at 5701.2 \AA\, the spatial equivalent width distribution does not follow the structure of the globular
cluster. Instead it shows a distinct region with an equivalent width 
 that is about 50\% higher than for the rest of the field of view.}
\section{Conclusions}
Measuring equivalent widths in $\sim$ 10,000 lines of sight only became feasible with an automatic approach to extracting and modeling
stellar spectra as presented in Paper I.
Our equivalent width measurements of DIBs in the composite spectrum of thousands of stars reach a very high accuracy down to few m$\AA$.
{ The measured universal ratios are in perfect agreement with results based on high-resolution spectrographs.}
We evidently resolve small structures in the order of milliparsec and can confirm the correlation of numerous DIBs on small
scales within a physically connected region. We demonstrate the feasibility of this approach and outline
the main challenges involved.  

\begin{itemize}

\item The \NaI\, doublet shows some filament-like structure in the FOV of $300' \times 
300'$ on the sky against NGC\,6397 .

\item We mapped a clumpy structure of \KI with a projected size of $\approx 4,000$ AU assuming a distance of 100 pc.

{ 
\item We were able to trace  optical depth changes of the absorber across the field of view via line ratios of the doublets of \NaI\, and \KI\, (and more indirectly 
from the DIB 5780/5797 ratio).}

\item The projected spatial distribution of DIB 5705 observed in the mapped equivalent
 widths in Fig. \ref{Fig:highres_ism}
stands out as being particularly inhomogeneous.
 While the overall equivalent width of DIB 5705 is rather low with $\approx 40$ m\AA\, a value above 60 m\AA\, at the bottom
right extends across several independent bins. This could indicate an independent carrier of DIB 5705. It remains an exciting species for our analysis of further GC sight lines.

\item A  puzzling finding is the anti-correlation of DIB 6283 with several DIBs. It is the strongest 
against DIB 5797 and \KI. 
We have rejected several systematic errors in the sky modeling, correlations with the environmental conditions of the observations and
reduction issues as possible cause of this effect.{  The findings support the possibility
of the carrier of DIB 6283 favoring regions of ionized gas. 
A future study based on more targets will have to evaluate whether the behavior of DIB 6283 is commonly observed and whether the detected ratios of DIB 5780/5797 which deviate from the universal trend is typical on local scales.  }

\end{itemize}
We currently target 25 GCs with VLT/MUSE. Most of them will be observed in several epochs which will give another boost in S/N as well as constraints on 
temporal variability and tangential motions.
We open a new window on studying DIBs spatially resolved at very small scales. This initial pilot study demonstrates the feasibility
of the instrument to trace ISM structures despite its comparably low spectral resolution.
\begin{acknowledgements}
Many thanks to Dan Welty and Scott Friedman who generously provided data of DIB and ISM measurements in private communication. { We also thank the referee for the thorough reports improving the flow of the paper.}
\\
AMI acknowledges support from the Spanish MINECO through project AYA2015-68217-P and the Agence Nationale de la Recherche through the STILISM project
(ANR-12-BS05-0016-02).
\\
This work has made use of data from the European Space Agency (ESA)
mission {\it Gaia} (\url{https://www.cosmos.esa.int/gaia}), processed by
the {\it Gaia} Data Processing and Analysis Consortium (DPAC,
\url{https://www.cosmos.esa.int/web/gaia/dpac/consortium}).

\end{acknowledgements}

\appendix
\label{sec:appendix}
\begin{appendix}
\section{}
\subsection{Modeling of absorption features}
\label{sec:model}
{
As briefly described in Section \ref{sec:MUSE integrated spectrum}, all absorption features were modeled 
with the evolutionary algorithm described in \citet{quast05} and applied for molecular species in \citet{wendt2012,wendt2014}.
The \KI and \NaI\, features were fitted as Voigt profiles together with a low order polynomial for the  local continuum
and taking the appropriate instrument resolution into account.
For  the \KI doublet and \NaD doublet both transitions of each doublet shared the same broadening parameter.
In the high resolution UVES spectrum multiple velocity components were resolved for the \NaD doublet and fitted simultaneously at
a tied velocity offset. The wavelength range of \KI was not covered in the available UVES the data.

The broad DIB features  have no complex line profile at the resolution of these data. Many are known to show asymmetric shapes,
even though that is barely resolved by MUSE. This was taken into account by fitting regions of DIBs with 
usually two Gaussians to each observed DIB feature simultaneously.
The equivalent widths were derived by integrating over the modeled Gaussians per DIB
individually to account for possible overlap regions.
}
\subsection{Complementary plots and maps}
\label{sec:highmap}

\begin{figure}[]
   \centering
\includegraphics[width=0.4\textwidth]{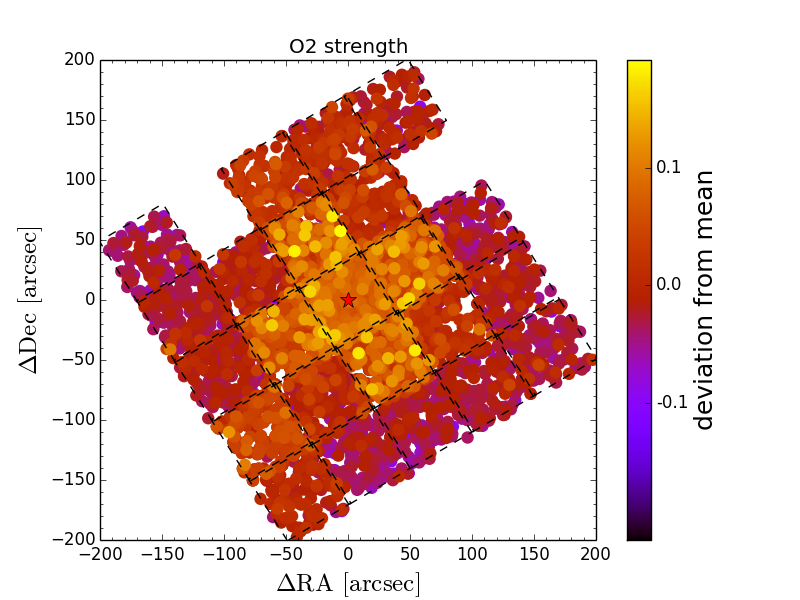}
\includegraphics[width=0.4\textwidth]{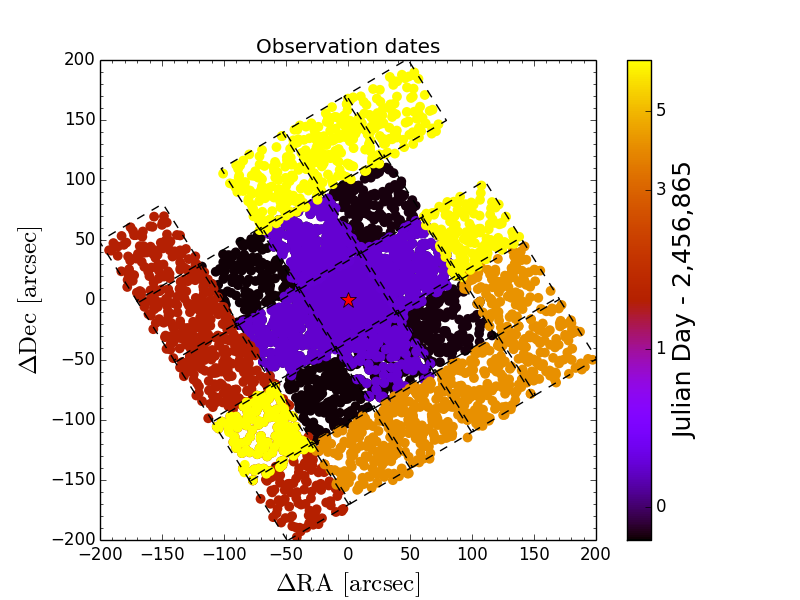}\\
      \caption{{\it top:} Strengths of \Otwo\, in the sky model per individual star with a S/N above 20. 
        {\it bottom:} The day of observation relative to JD 2,456,865. The separate pointings are framed in black dashed lines.
       We note, that several pointings overlap and only the latest visually contributes to this plot.}
         \label{Fig:o2}
   \end{figure}

   \begin{figure*}[]
   \centering
   \includegraphics[width=0.45\textwidth]{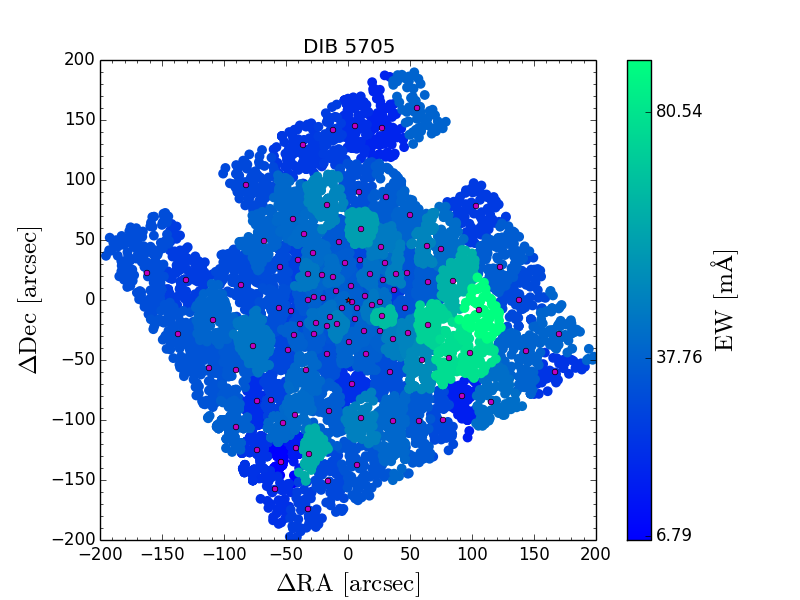}
  \includegraphics[width=0.45\textwidth]{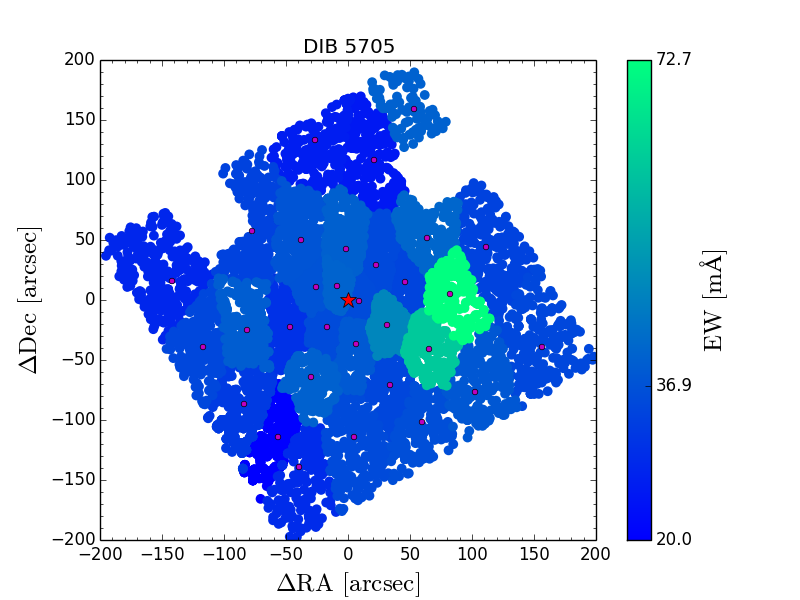}\\
      \includegraphics[width=0.45\textwidth]{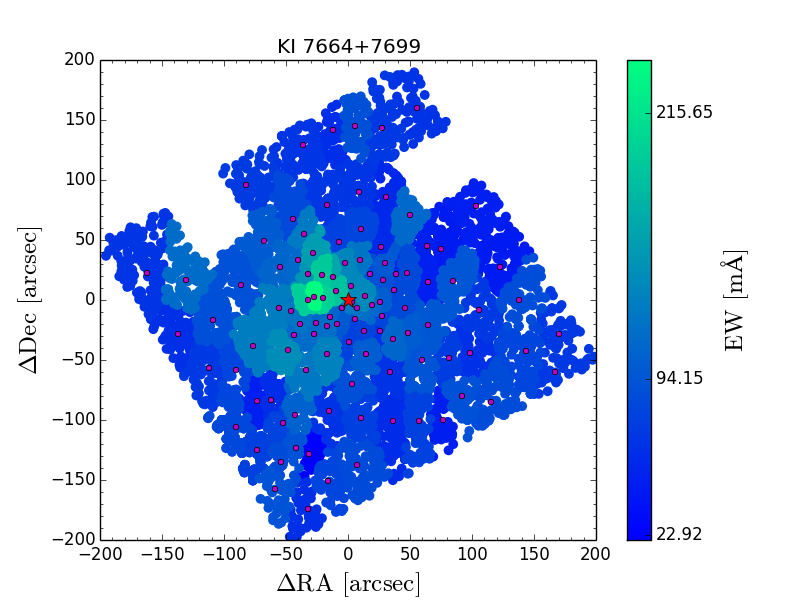}
\includegraphics[width=0.45\textwidth]{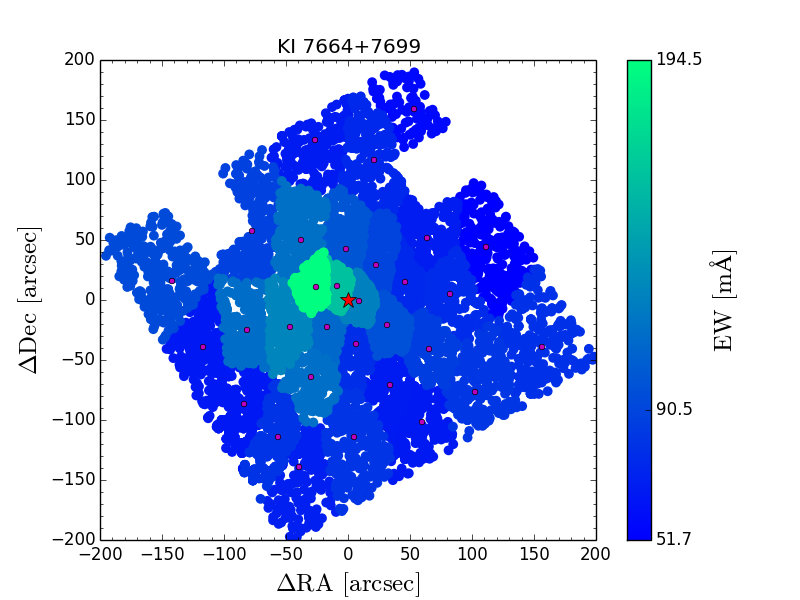}\\
      \caption{{{\it top:} Equivalent width maps for DIB 5705. On the {\it left} with 107 bins, on the {\it right} with 31 bins. 
        {\it bottom:}for the \KI doublet. On the {\it left} with 107 bins, on the {\it right} with 31 bins. The magenta circles 
        mark the individual Voronoi bins.
        The red star denotes the cluster's center.
        The uncertainty is in the order of 14 m\AA\, for the high resolution map and 8 m\AA\, for the 31 bins on average.}}
         \label{Fig:highres_ism}
   \end{figure*}
   \begin{figure*}[]
   \centering
   \includegraphics[width=0.45\textwidth]{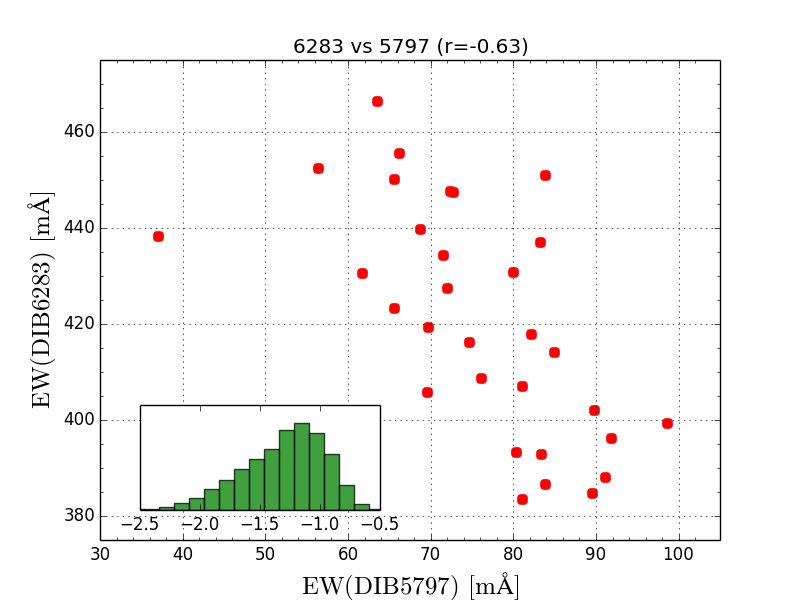}
   \includegraphics[width=0.45\textwidth]{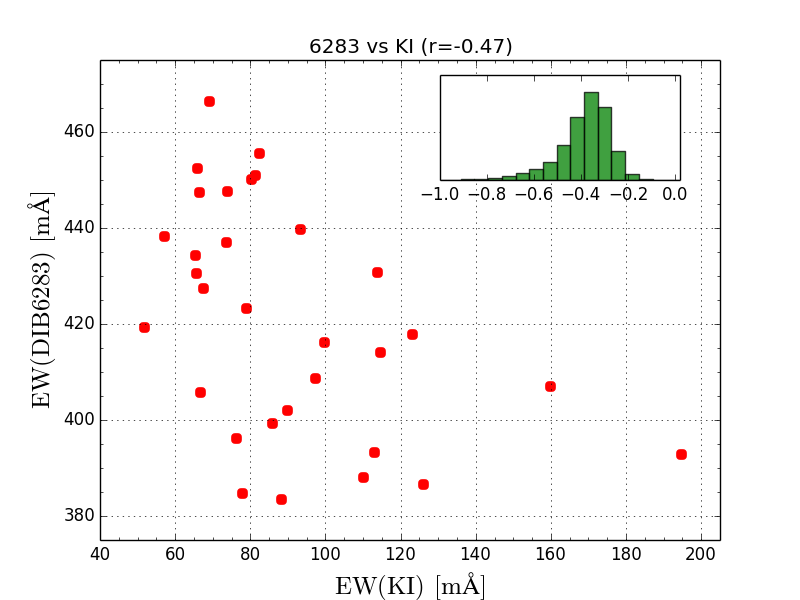}
    \caption{Equivalent widths of DIB 6283 vs DIB 5797 and \KI.}
         \label{Fig5797ratio}
   \end{figure*}
   \begin{figure*}[]
   \centering
   \includegraphics[width=0.45\textwidth]{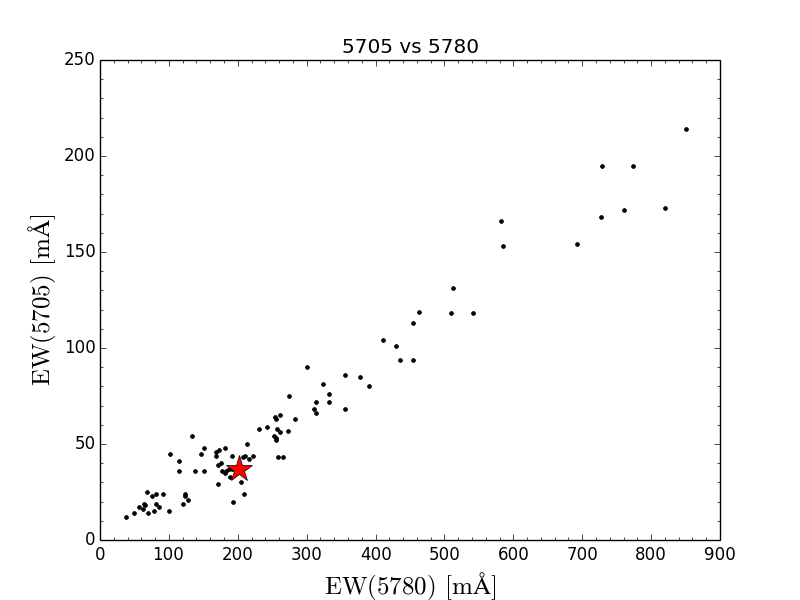} 
   \includegraphics[width=0.45\textwidth]{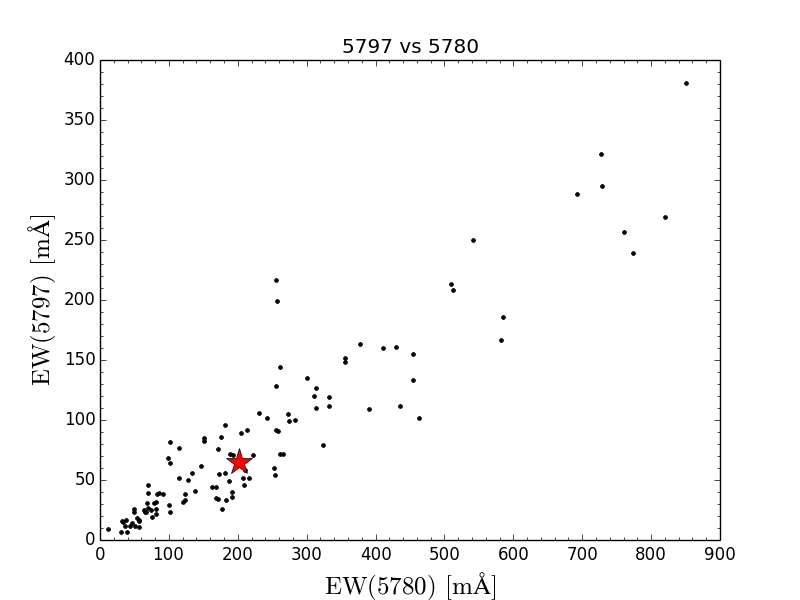}\\
   \includegraphics[width=0.45\textwidth]{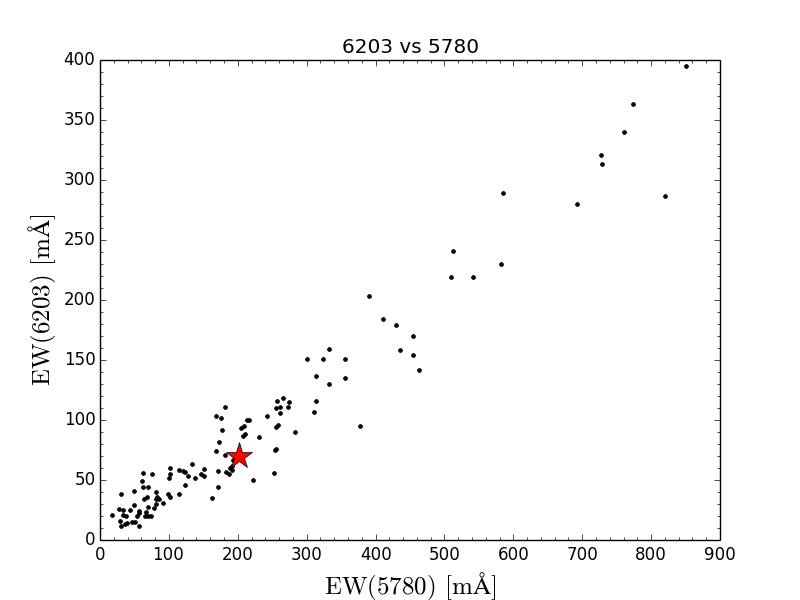}
   \includegraphics[width=0.45\textwidth]{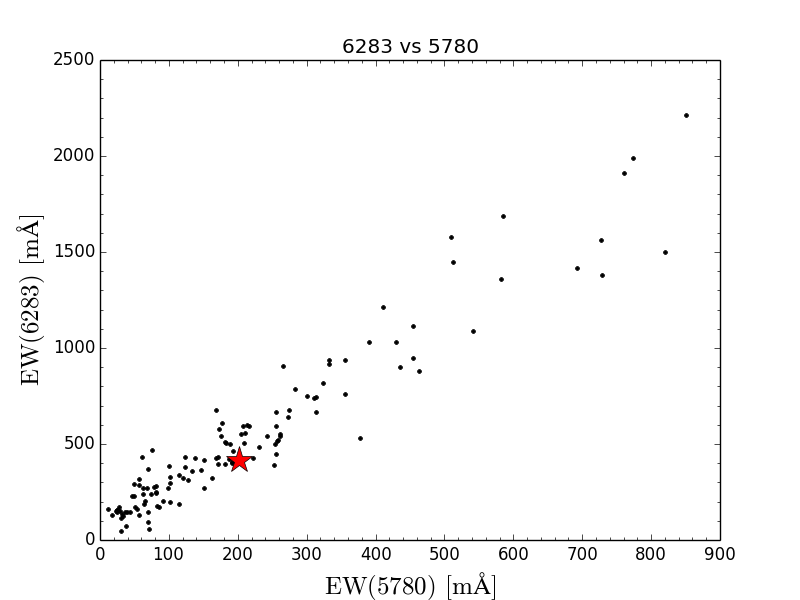}\\
   \includegraphics[width=0.42\textwidth]{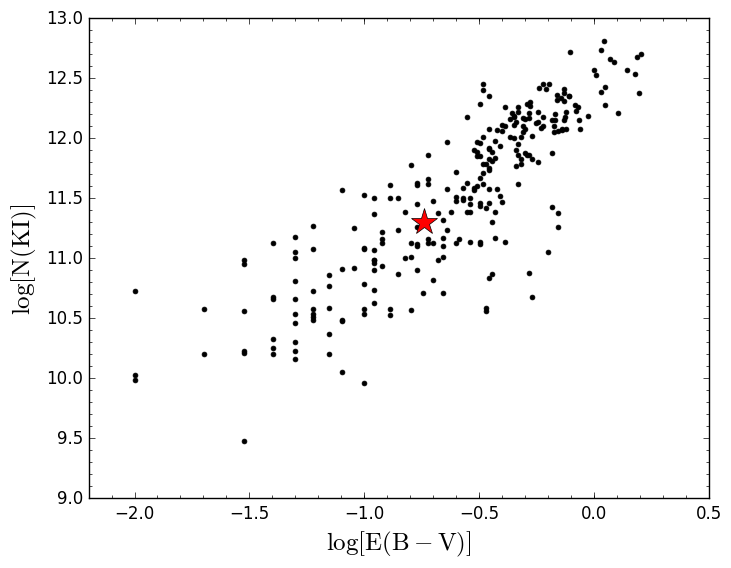}
    \caption{Equivalent widths of DIB 5705, 5797, 6203, 6283 vs DIB 5780 and N(\KI) vs E(B-V). The black points are data
    from \citet{Friedman11}, the red star is the measurement based on the composite spectrum. }
        \label{fig:friedman}
   \end{figure*}
\end{appendix}
\end{document}